\documentclass{aa}  
\usepackage{rotating}
\usepackage{graphicx}
\usepackage{txfonts}
\usepackage{lscape}
\usepackage[bf,nooneline]{subfigure}

\bibpunct{(}{)}{;}{a}{}{,}

\begin{document}

   \title{The population of early-type galaxies: how it evolves with time and how it differs from passive and late-type galaxies} 
                                                     
   \author{S. Tamburri\inst{1,2}\thanks{E-mail: 
sonia.tamburri@brera.inaf.it}, P. Saracco$^{1}$, M. Longhetti$^{1}$, A. Gargiulo$^{1}$, I. Lonoce$^{1,2}$, F. Ciocca$^{1,2}$}

   \institute{INAF - Osservatorio Astronomico di Brera, Via Brera 28, 20121 Milano,
Italy \and
Dipartimento di Scienza e Alta Tecnologia, Universit\'a degli Studi dell'Insubria, via Valleggio 11, 22100 Como, Italy.}

   \date{Received 2014 April 22 / Accepted 2014 September 2}

\abstract{}{There are two aims to our analysis. On the one hand we are interested in addressing whether a sample of morphologically selected early-type galaxies (ETGs) differs from a sample of passive galaxies in terms of galaxy statistics. On the other hand we study how the relative abundance of galaxies, the number density, and, the stellar mass density for different morphological types change over the redshift range $0.6\leq z\leq2.5$.}
{From the 1302 galaxies brighter than Ks(AB)=22 selected from the GOODS-MUSIC catalogue, we classified the ETGs, i.e. elliptical (E) and spheroidal galaxies (E/S0), on the basis of their morphology and the passive galaxies on the basis of their specific star formation rate (sSFR$\leq$10$^{-11}$ yr$^{-1}$). Since the definition of a passive galaxy depends on the model parameters assumed to fit the spectral energy distribution of the galaxy, in addition to the assumed sSFR threshold, we probed the dependence of this definition and selection on the stellar initial mass function (IMF).} {We find that spheroidal galaxies cannot be distinguished from the other morphological classes on the basis of their low star formation rate, irrespective of the IMF adopted in the models. In particular, we find that a large fraction of passive galaxies ($>30$ \%) are disc-shaped objects and that the passive selection misses a significant fraction ($\sim26$ \%) of morphologically classified ETGs. 
Using the sample of 1302 galaxies morphologically classified into spheroidal galaxies (ETGs) and non-spheroidal galaxies (LTGs), we find that the fraction of these two morphological classes is constant over the redshift range $0.6\leq z\leq2.5$, being 20-30 \% the fraction of ETGs and 70-80 \% the fraction of LTGs. However, at $z<1$ these fractions change among the population of the most massive (M$_*\geq10^{11} M_{\odot}$) galaxies, with the fraction of massive ETGs rising up to 40 \% and the fraction of massive LTGs decreasing to 60 \%. Parallel to this trend, we find that the number density and the stellar mass density of the whole population of massive galaxies increase by almost a factor of $\sim 10$ between $0.6\leq z\leq2.5$, with a faster increase of these densities for the ETGs than for the LTGs. Finally, we find that the number density of the highest-mass galaxies both ETGs and LTGs (M$_*>3-4\times10^{11} M_{\odot}$) does not increase from $z\sim2.5$, contrary to the lower mass galaxies. This suggests that the most massive galaxies formed at $z>2.5-3$ and that the assembly of such high-mass galaxies is not effective at lower redshift.}{}


\titlerunning {The population of ETGs: how it evolves with time and how it differs from passive and late-type galaxies}
\authorrunning {Tamburri et al. }

\maketitle

\section{Introduction}
Understanding how galaxies formed and assembled their mass along the cosmic time is a critical and not completely resolved issue of modern observational cosmology. Many studies have focused their attention on the population of early-type galaxies (ETGs), i.e. elliptical (E) and spheroidal galaxies (E/S0), since these galaxies contain most of the stellar mass and host the most massive galaxies in the local Universe (e.g. Baldry et al. \citeyear{baldry04}, Fukugita et al. \citeyear{fukugita98}). 
For these reasons they are thought to be of fundamental importance in order to trace the history of stellar mass assembly and to explain the physical processes driving galaxy evolution.
Early-type galaxies in the local Universe form an almost homogeneous family of systems mainly characterised by negligible star formation, relatively old stellar populations, red colours (e.g. Bernardi et al. \citeyear{bernardi03}, Faber et al. \citeyear{faber07}), and tight correlations among observed quantities such as luminosity, colour, size, and velocity dispersion \citep[][and references therein]{renzini06}. Despite their apparent simplicity, the physical processes involved in the formation and evolution of ETGs are still unclear and many attempts have been done to solve this issue starting from the analysis of the galaxy morphology. In fact, the different morphology of galaxies results from the different processes with which their stellar mass has been assembled.

If visual morphological classification at low redshift is not challenging, at higher redshift morphological classification was not always feasible because of observational limits. Moreover, the increasing sizes of galaxy samples has favoured the development and the use of classification methods faster than the visual and often based on the photometric properties of local galaxies. Following this, multi-colour classifications were adopted to split red galaxies with an old stellar population, in principle ETGs, from the complementary population of blue star forming galaxies, which should be predominantly discs (e.g. Strateva et al. \citeyear{strateva01}, Bell et al. \citeyear{bell04}, Ilbert et al. \citeyear{ilbert10}). However, a rigid galaxy classification starting from the bimodality of the colour-magnitude diagram cannot take into account events that alter the galaxy colours, such as dust extinction of star forming galaxies \citep{williams09} or quenched star formation in late-type objects \citep{bell08}. Another technique aimed at distinguishing ETGs from late-type galaxies (LTGs) makes use of to the star formation activity of the galaxies, dividing passive and active galaxies according to the efficiency of their star formation rate with respect to their mass (e.g. Feulner et al. \citeyear{feulner05}, Drory et al. \citeyear{drory05}, Fontana et al. \citeyear{fontana09}, Cassata et al. \citeyear{cassata11}).
Alternatively, the shape parameter $n$ related to the S\'{e}rsic profile has been also used to separate disc-like surface brightness profiles ($n<2-2.5$) from elliptical ones ($n>2-2.5$) (e.g. Shen et al. \citeyear{shen03}, Cassata et al. \citeyear{cassata05}, Buitrago et al. \citeyear{buitrago13}). However, since there is not a one-to-one correspondence between the photometric properties of a galaxy and its shape, all these methods fail to distinguish the morphology of galaxies that produce mixed samples of ETGs and discs. Moreover, the photometric properties of the ETGs observed in the local Universe are not necessarily those characterising the ETGs at high-z.

To disentangle the evolutionary phases of spheroids and/or discs it is important to separate these two morphological classes, since their different dynamic and shape suggest a different mass assembly history. 
In the current hierarchical scenario, ETGs should appear later in time as the product of mergers between small subunits of disc-like or irregular galaxies, following a hierarchical build-up similar to that of their host dark matter halos \citep{delucia06}. Thus, to have insight into how the matter aggregates in the galaxies we see today for each morphological type, it is important to keep progenitors separate from descendants. This way we can trace the evolutionary path followed by the two classes of objects, analysing the mean properties of each morphological type and constraining the mechanisms characterising their growing and shaping.

Up to now, there is still an ongoing debate on which population dominates the Hubble diagram at high redshift and how the fraction of galaxies populating the two main morphological categories evolves with time.
Some studies find evidence that in the distant Universe irregular/peculiar galaxies were the predominant morphological class and that visually classified disc galaxies were a small fraction of the entire population, while a large ($\sim40$ \%) population of ellipticals was already present at $z>2$ (Mortlock et al. \citeyear{mortlock13}, Conselice et al. \citeyear{conselice11}). Other studies on massive galaxies find changes by a factor of three in the fraction of elliptical and lenticular galaxies between $z\sim0$ and $z\sim3$ and a constant amount of pure discs in the same redshift range \citep{buitrago13}. Conversely, several analyses find that most of the galaxies at high redshift cannot be classified within the Hubble scheme (van den Bergh et al. \citeyear{vandenbergh01}, Papovich et al. \citeyear{papovich05}; Cameron et al. \citeyear{cameron11}). Various physical mechanisms have been suggested as the drivers of the evolution of the shape of galaxies, many of them describing a scenario in which blue star forming objects turn into red-bulge dominated systems with quenched star formation. 
The morphological transformation has been ascribed to several processes such as major and minor galaxy mergers (Toomre \& Toomre \citeyear{toomre72}, White \& Rees \citeyear{white78}, Fall \citeyear{fall79}, Naab \& Burkert \citeyear{naab03}, Bournaud et al. \citeyear{bournaud05}, \citeyear{bournaud07}), gravitational fragmentation of gas-rich discs into big clumps with a subsequent migration into a central bulge (van den Bergh et al. \citeyear{vandenbergh96}, Genzel et al. \citeyear{genzel08}, Bournaud et al.\citeyear{bournaud08}), and continuous gas supply via cold streams in disc galaxies that drives the disc instability and the subsequent growth of a spheroid \citep{dekel09b}.
However, all these mechanisms need to be reconciled with the number of galaxies of a given mass and shape observed in the local Universe with those detected at high redshift, where observational bias and effects of misclassification of galaxy morphology can easily occur.

In this paper we select a sample of early-type galaxies classified on the basis of morphology and a sample of passive galaxies classified on the basis of their specific star formation rates, i.e the star formation rate (SFR) normalised for the stellar mass. There are two aims to our analysis. On one hand we are interested in addressing whether the two selection criteria select approximately the same galaxies, i.e. elliptical and spheroidal galaxies, and hence they can be used interchangeably or instead they produce samples significantly different whose properties cannot be compared. Moreover, we would like to test the dependence  of the selection criterion based on the specific SFR on the different model assumptions, in particular on the choice of the stellar initial mass function (IMF). On the other hand, we are interested in describing how the composition of the population of galaxies is changed with time both in terms of relative abundance of different morphological types and in terms of their numerical growth.

The structure of the paper is as follow. In Section 2 we first describe the main properties of the GOODS-MUSIC catalog, the data set on which our analysis is based. Then, we describe the selection of our sample of galaxies, their morphological classifications, and the methodology used to derive their main physical properties, i.e. stellar mass, age, effective radius, and axial ratio. Finally, we describe the derivation of their specific star formation rate and the criteria used to define passive and non-passive galaxies. Section 3 presents the differences between the morphological sample and the passive sample. Section 4 describes the dependence of the passive galaxy selection on the IMF adopted in the models. In Section 5 we present a study of the evolution of the relative fraction of galaxies, of the number density of galaxies, and of the stellar mass density for different morphological types over the redshift range $0.6\leq z \leq2.5$. In section 6 we summarise the results and present our conclusions.

Throughout this paper, we use a standard cosmology with H$_0=70$ Km s$^{-1}$ Mpc$^{-1}$, $\Omega_m=0.3$ and, $\Omega_{\Lambda}=0.7$. All the magnitudes are provided in the AB system.

\section{Sample selections and parameter estimates}

\begin{figure*}
\includegraphics[width=18 truecm]{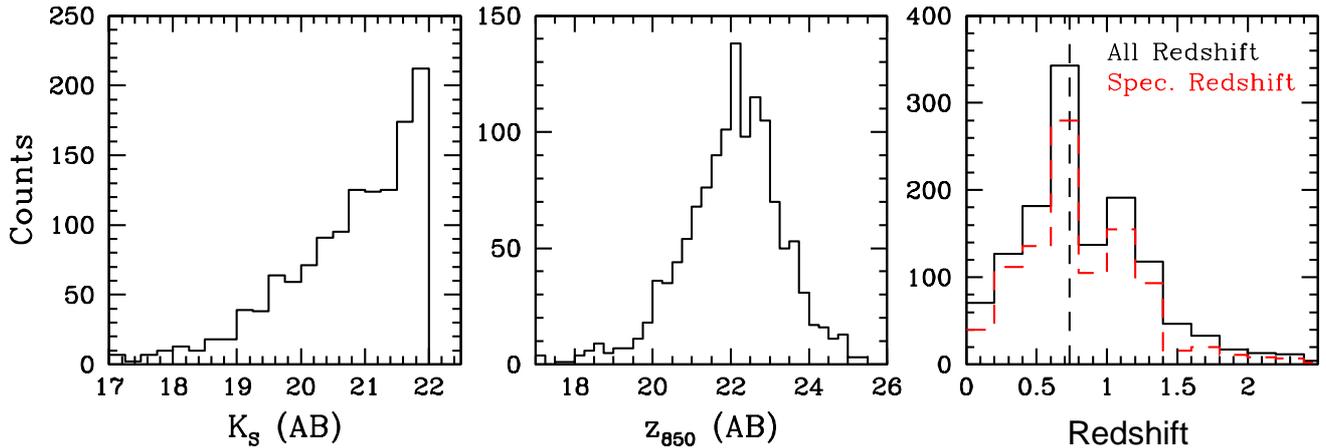}\\
\caption{\textit{Left and centre:} apparent magnitude distributions of the sample of 1302 galaxies in the Ks and $z_{850}$ bands. \textit{Right:} redshift distribution of the Ks$\leq$22 sample. The continuous black line and the dashed red line indicate the whole sample and the objects with spectroscopic redshift, respectively. The vertical dotted line at z$\sim$0.74 indicates the median of redshift distribution.}
\label{fig:zdistr}
\end{figure*}

The analysis presented in this paper is based on a magnitude limited sample extracted from the GOODS-MUSIC catalogue (Grazian et al. \citeyear{grazian06}, Santini et al. \citeyear{santini09}) of the Great Observatory Origins Deep Survey \citep[GOODS-South; ][]{giavalisco04}. The MUSIC catalogue is entirely based on public data collected on the $\sim$143 arcmin$^2$ of the GOODS-South field and includes photometric and spectroscopic redshift measurements and a multi-wavelength photometry in 15 bands: three U bands; four HST-ACS filters (\textit{B}$_{435}$, \textit{V}$_{606}$, \textit{i}$_{775}$, \textit{z}$_{850}$); three VLT-ISAAC \textit{J}, \textit{H} and, \textit{Ks} bands; and five Spitzer bands at 3.5, 4.5, 5.8, 8.0, and 24.0 $\mu$m.
The spectroscopic data come from the public ESO-VLT spectroscopic survey (Vanzella et al. \citeyear{vanzella08}, Popesso et al. \citeyear{popesso09}, Balestra et al. \citeyear{balestra10}) and other spectroscopic campaigns.

In order to have a good spectroscopic redshift coverage and to perform an accurate morphological classification of galaxies up to z$\sim$2.5, we selected from the MUSIC catalogue the 1302 galaxies brighter than \textit{Ks}=22. The distributions of the galaxies as a function of the apparent magnitude in the Ks and $z_{850}$ bands are shown in the left and central panels of Fig. \ref{fig:zdistr}. The $z_{850}$ distribution is presented since the morphological classification of galaxies was done in this HST-ACS band. The central panel of Fig. \ref{fig:zdistr} shows that only 30 galaxies of the sample are not as bright as F850LP(AB)=24.5, the magnitude at which the morphological classification has been proved to be feasible and reliable (e.g. Mei et al. \citeyear{mei06}, Raichoor et al. \citeyear{raichoor12}, \citeyear{raichoor11}, Saracco et al. \citeyear{saracco11}).
The spectroscopic redshift completeness of this sample is of $\sim$76 \% and its redshift distribution is shown in the right panel of Fig. \ref{fig:zdistr}, where the total sample of 1302 galaxies (in black) and the galaxies with spectroscopic redshift (in red) are displayed.

The first aim of this work is to understand whether a sample of passive galaxies differs from a sample of morphologically selected ETGs, both in terms of selected galaxies and in terms of the main samples' properties. Starting from the magnitude limited sample of 1302 galaxies, we selected a sample of passive galaxies, defined as those galaxies with specific star formation rate sSFR$\leq$10$^{-11}$ yr$^{-1}$, and a sample of early-type galaxies, i.e. of galaxies visually classified as elliptical (E) and spheroidal (E/S0). In the next section we will adopt these definitions to distinguish passive galaxies from ETGs and we will show the selection methods in more detail.

\subsection{The morphological classification and the structural parameters}
\label{sec:morpho}
We built up a sample of morphologically classified galaxies that separates ETGs from all the remaining morphological classes. To this purpose we adopted a combination of two methods: a pure visual classification combined with a visual inspection of the residual images, obtained by fitting the galaxy surface brightness profiles with a S\'{e}rsic model.

\begin{figure*}
\centering
\subfigure
{\includegraphics[width=0.20\textwidth]{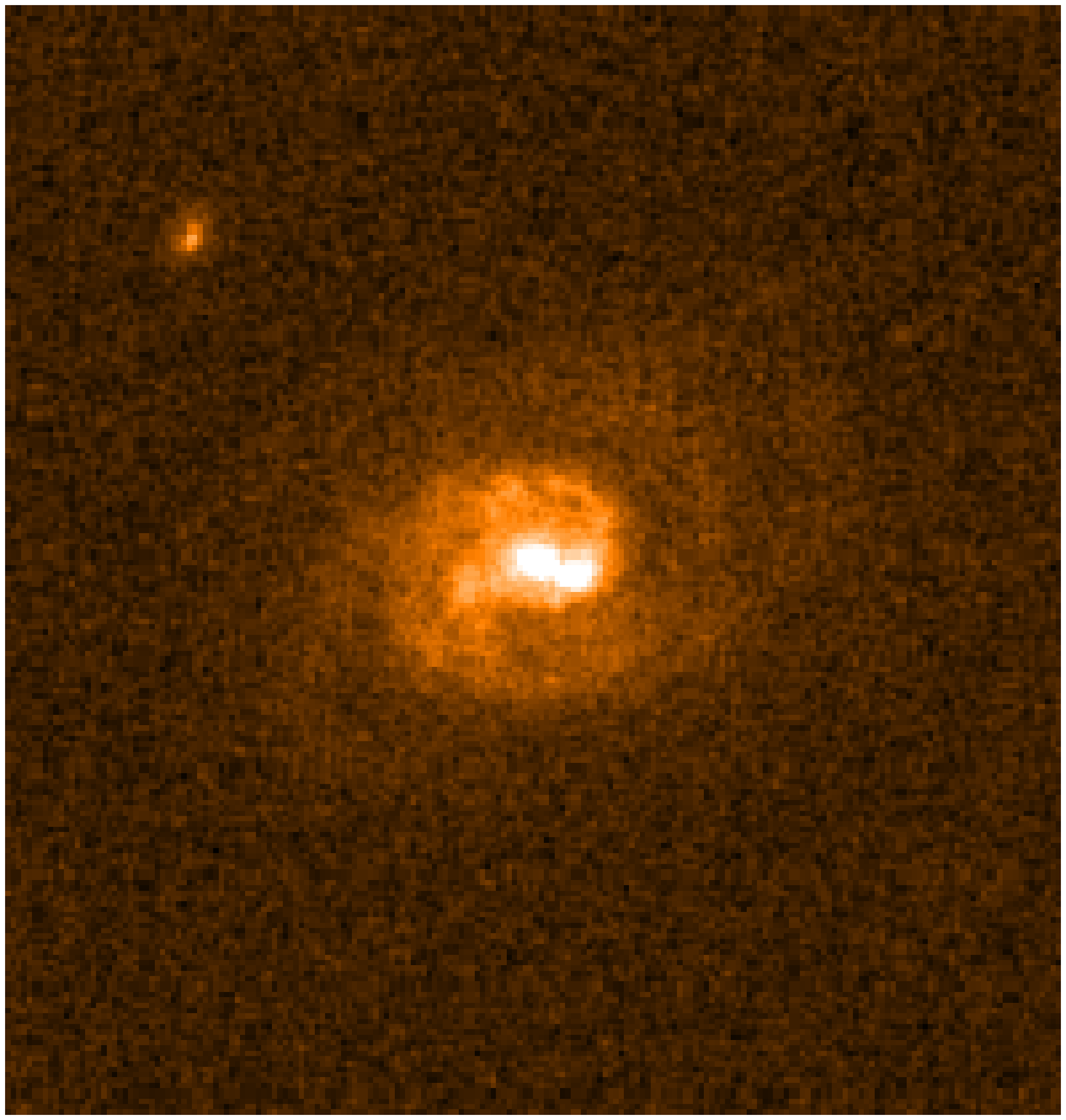}}
\quad
\subfigure
{\includegraphics[width=0.20\textwidth]{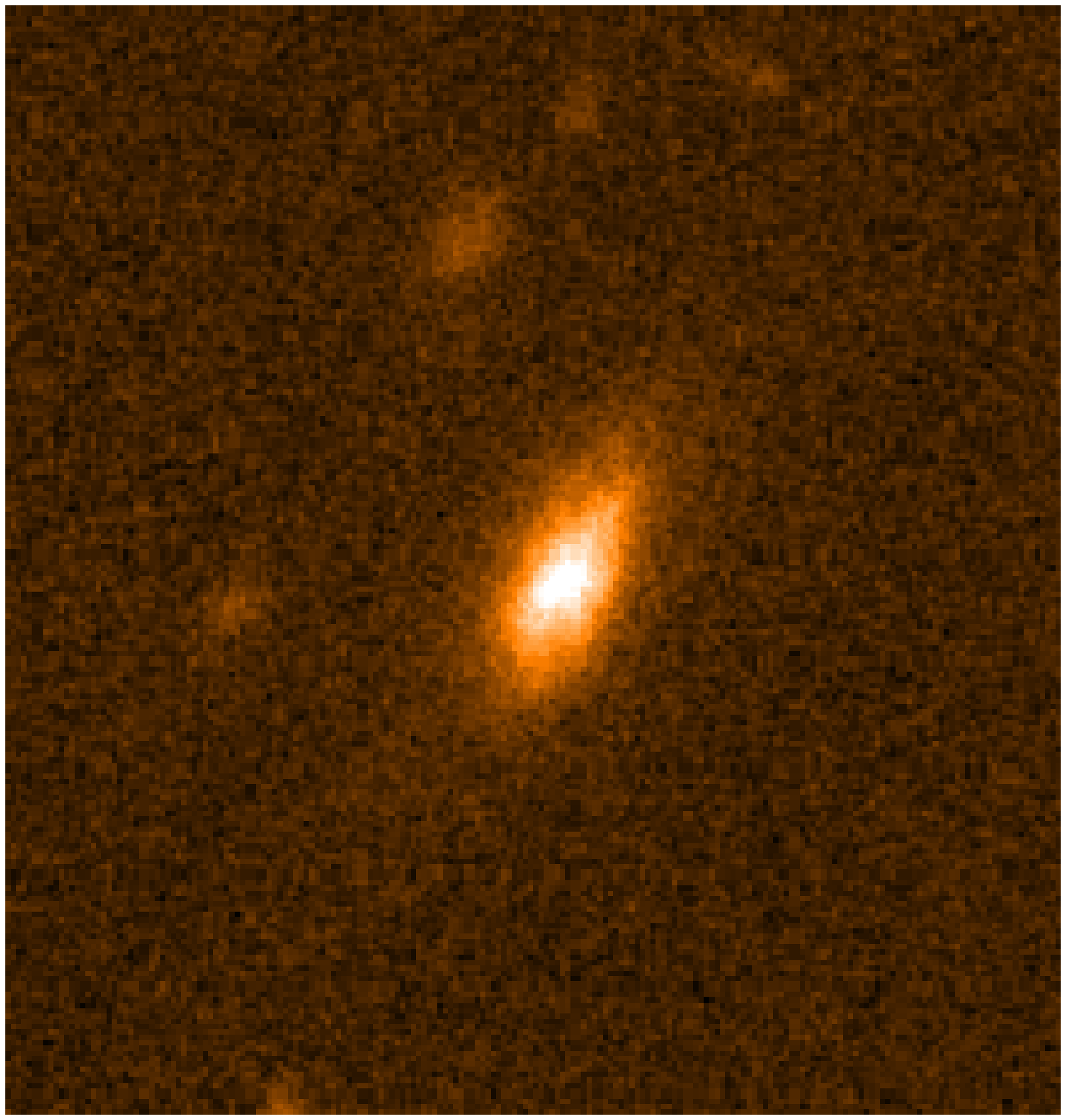}}
\quad
\subfigure
{\includegraphics[width=0.20\textwidth]{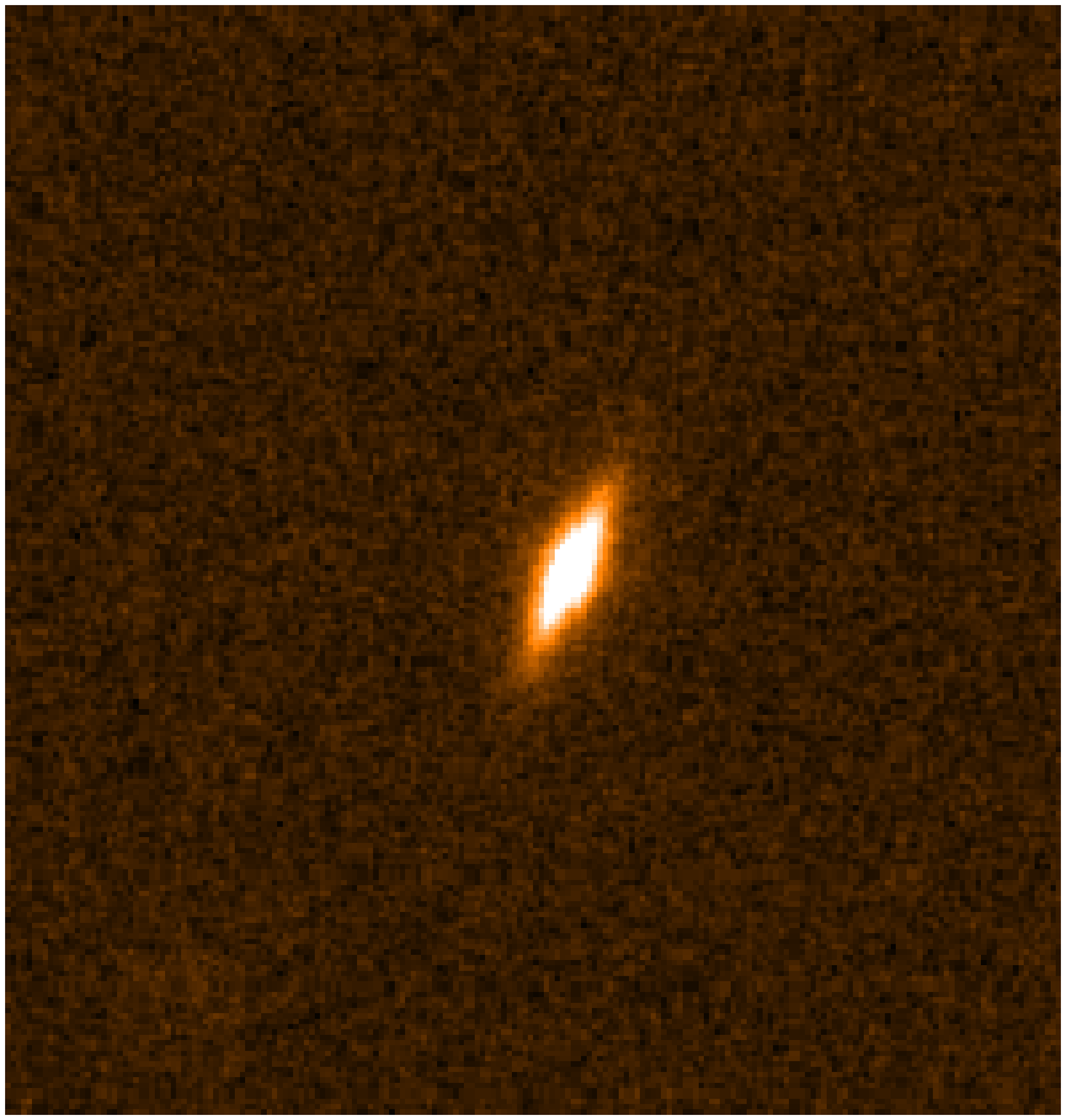}}
\quad
\subfigure
{\includegraphics[width=0.20\textwidth]{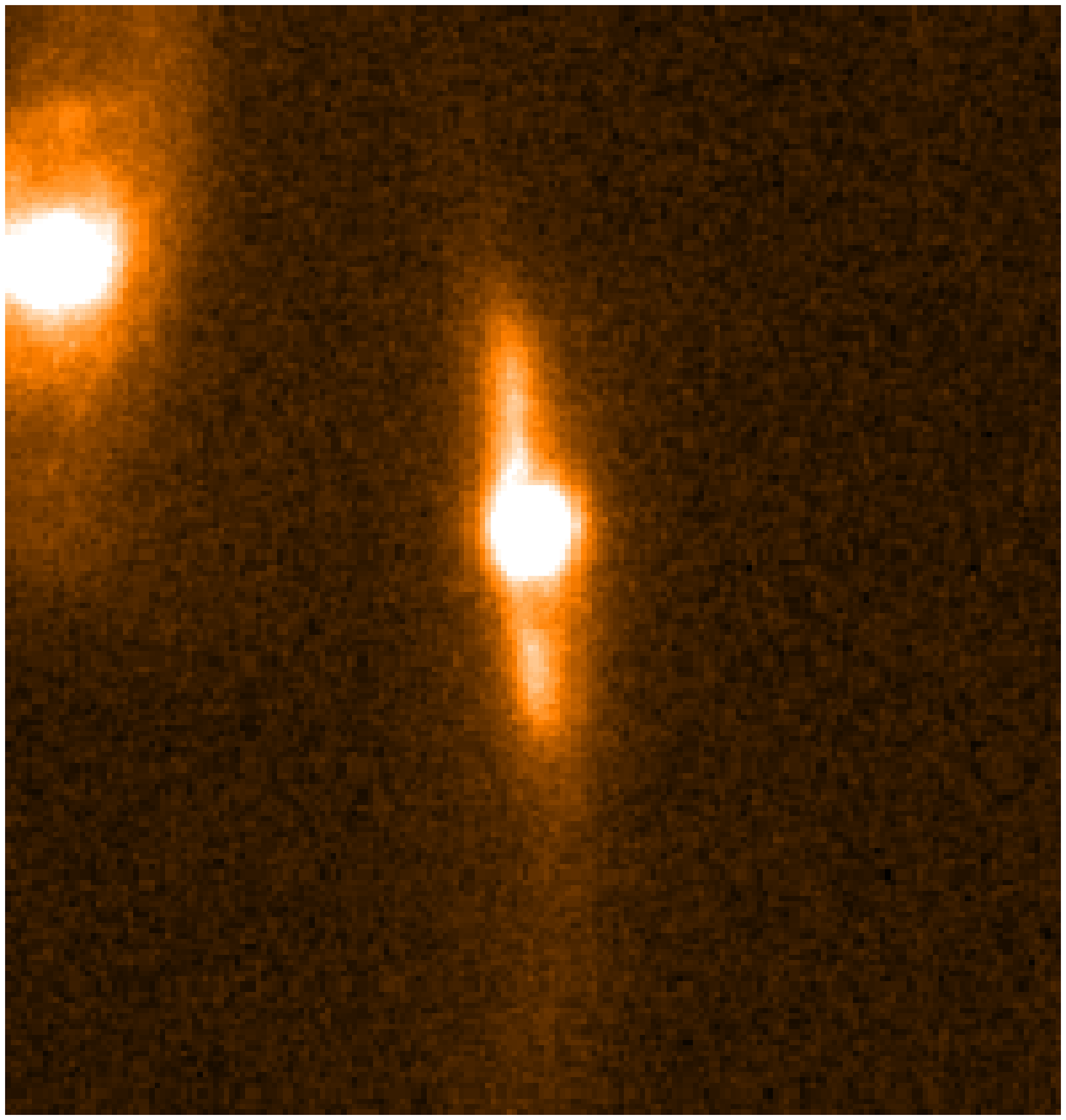}}
\\
\subfigure
{\includegraphics[width=0.20\textwidth]{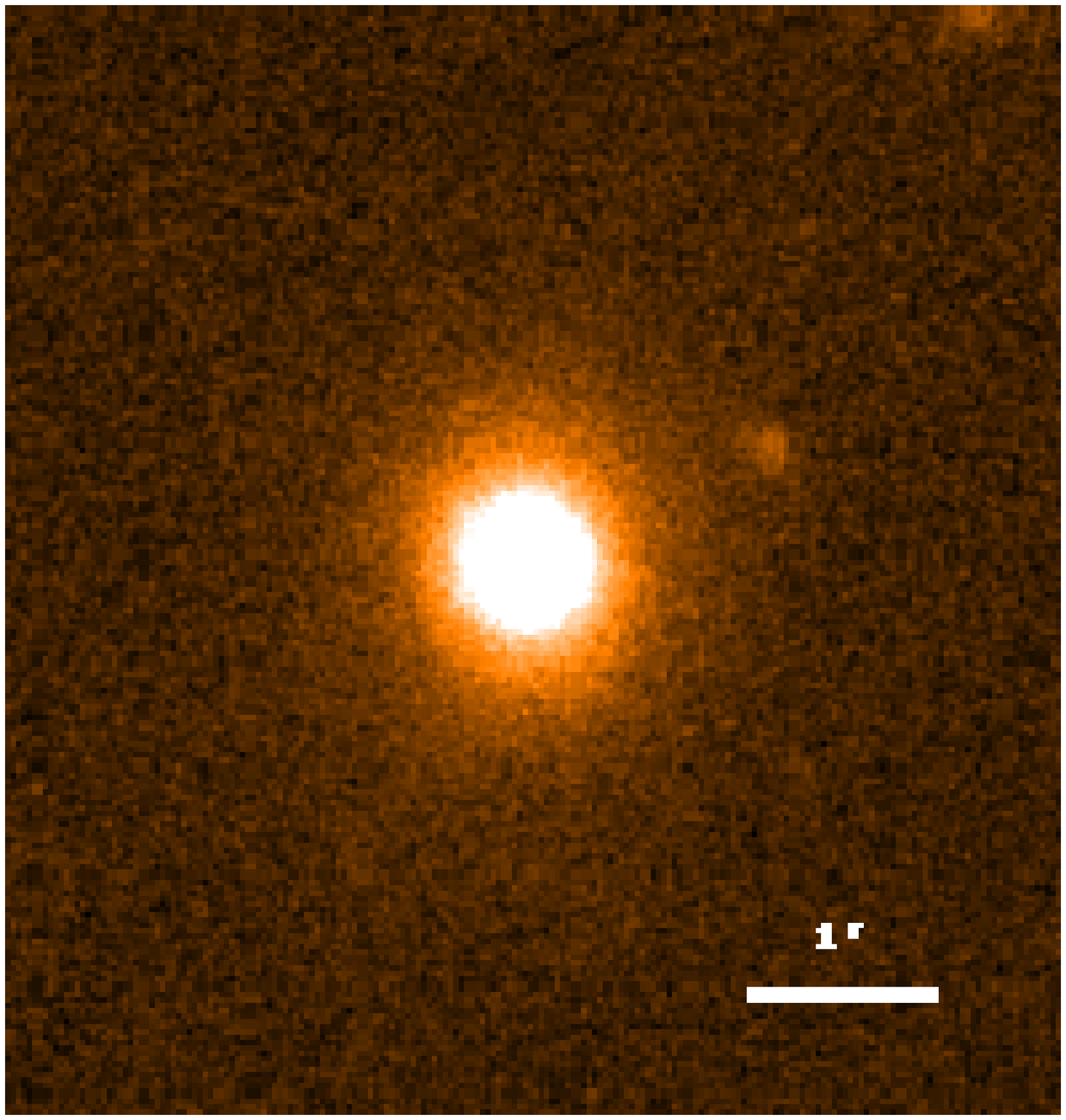}}
\quad
\subfigure
{\includegraphics[width=0.20\textwidth]{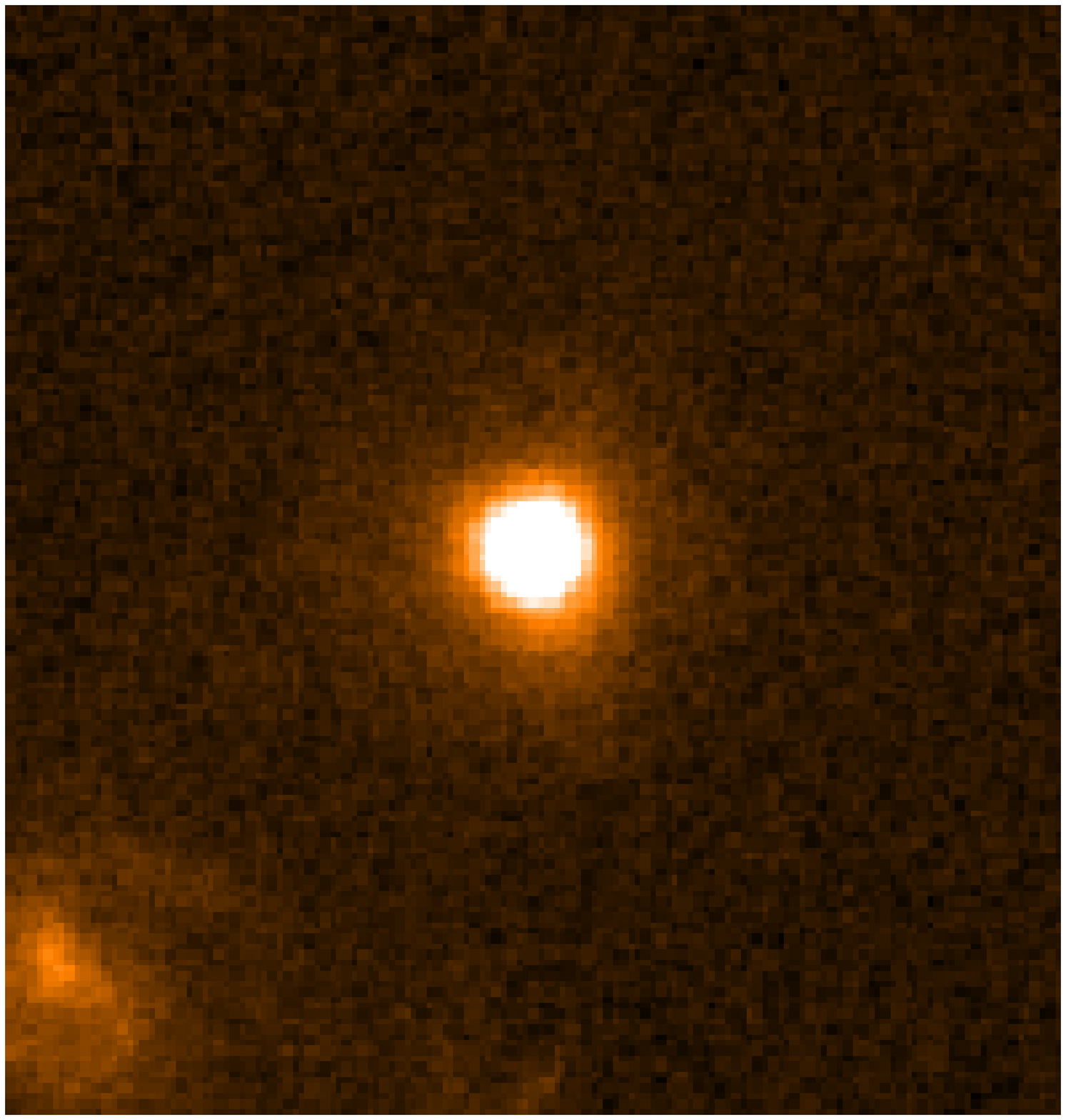}}
\quad
\subfigure
{\includegraphics[width=0.20\textwidth]{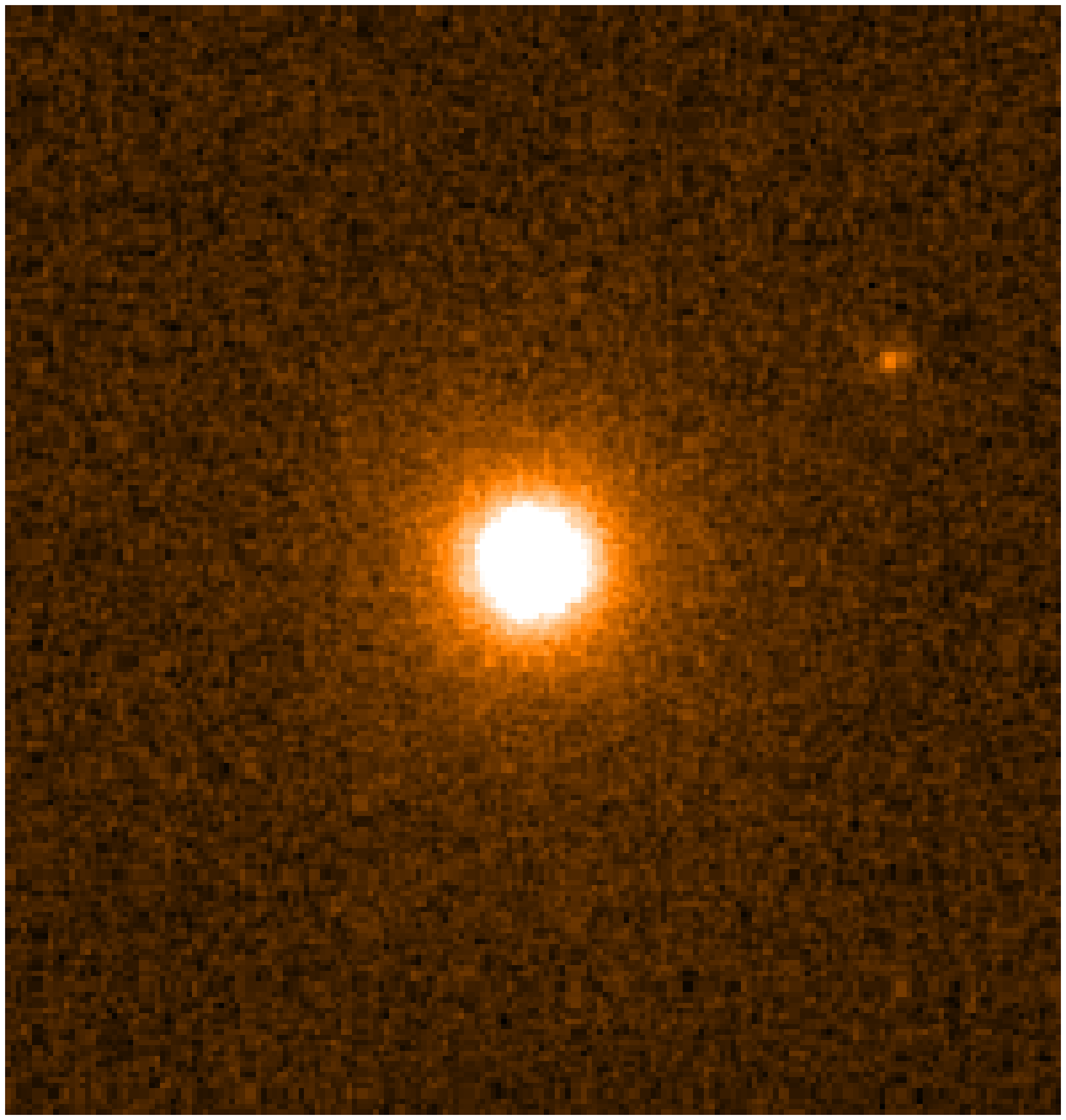}}
\quad
\subfigure
{\includegraphics[width=0.20\textwidth]{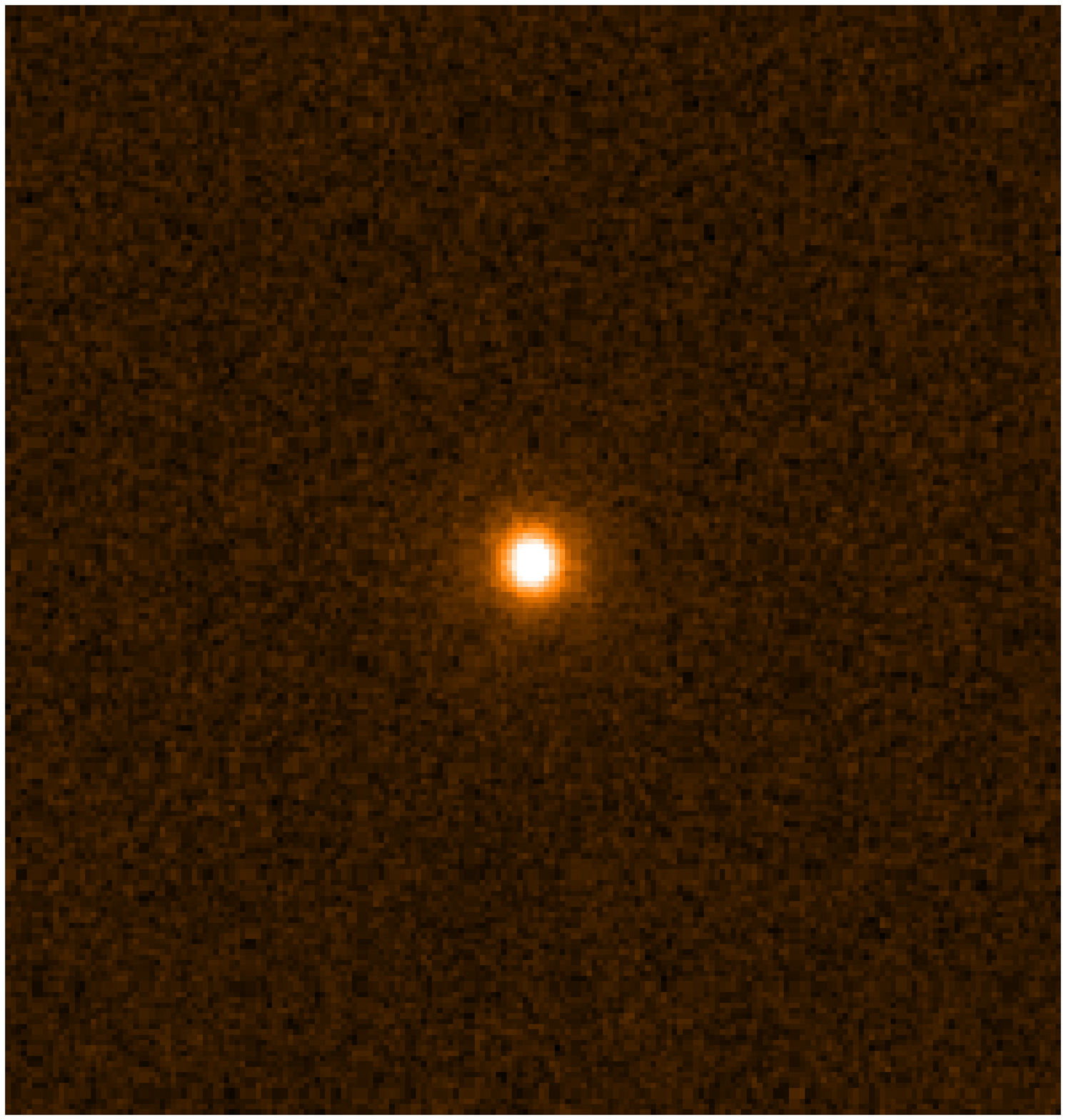}}
\caption{Examples of galaxies that fall into our late-type (\textit{top panels}) and early-type (\textit{bottom panels}) classification. The snapshots are taken from the F850LP-band images of the GOODS-S survey. }
\label{imm_galfit}
\end{figure*}

The visual morphological classification was independently performed on the high resolution HST/ACS F850LP images of the 1302 galaxies of the magnitude limited sample. We defined two morphological classes: 1) early-type galaxies, i.e. ellipticals (E) and transition E/S0 galaxies, which includes those centrally concentrated galaxies that look spheroid or ellipsoid with a regular shape and no hint of disc; 2) late-type galaxies, i.e. spirals (S), irregulars (Irr), and merging galaxies, i.e.  galaxies with a clear disc and/or irregular structure or with a hint of an ongoing merger process. 
The first step of the morphological classification was to identify galaxies with a clear presence of a disc and those with irregular shape, immediately classified as LTGs (e.g. Fig. \ref{imm_galfit}, top row). For the remaining galaxies we analysed both the images and the residual maps, the latter obtained by best fitting the HST-ACS/F850LP images with a 2D-PSF convoluted S\'{e}rsic model by means of the \texttt{GALFIT} software \citep[v. 3.0.4;][]{peng02}. This procedure gives the residual maps and the best-fitting parameters (S\'{e}rsic index \textit{n}, effective radius $r_e$, axis ratio) minimizing the differences between models and observations. The procedure adopted is described in Appendix A.

Our sample spans a wide redshift range $0.6\leq z\leq2.5$. Given this fact the F850LP filter samples different wavelength in the rest frame of the galaxies and as is known, the apparent structure of the galaxy can differ at different wavelengths (morphological K-corrections). In our case for $z\gtrsim1.2$ we do not expect a dependence of the morphology on the different rest-frame wavelengths sampled, since at this redshift the F850LP filter matches the U-band rest frame where the galaxy emission is mainly dominated by the light coming from the young stellar component. On the contrary, for $z<1$ the F850LP band samples the rest frame at $\lambda>4000$ \r{A} , dominated by the emission of the older stellar component. For these lower redshift galaxies we also performed the morphological classification in the F606W filter that at $z<1$ samples the U-band rest frame. We found that out of the almost 400 galaxies at $0.6<z<1$, only 2.5 \% have a different morphology in the two filters; for the other galaxies the morphological classification in the F606W and F850LP filters is the same.  

\begin{figure*}
\centering
\subfigure
{\includegraphics[width=0.20\textwidth]{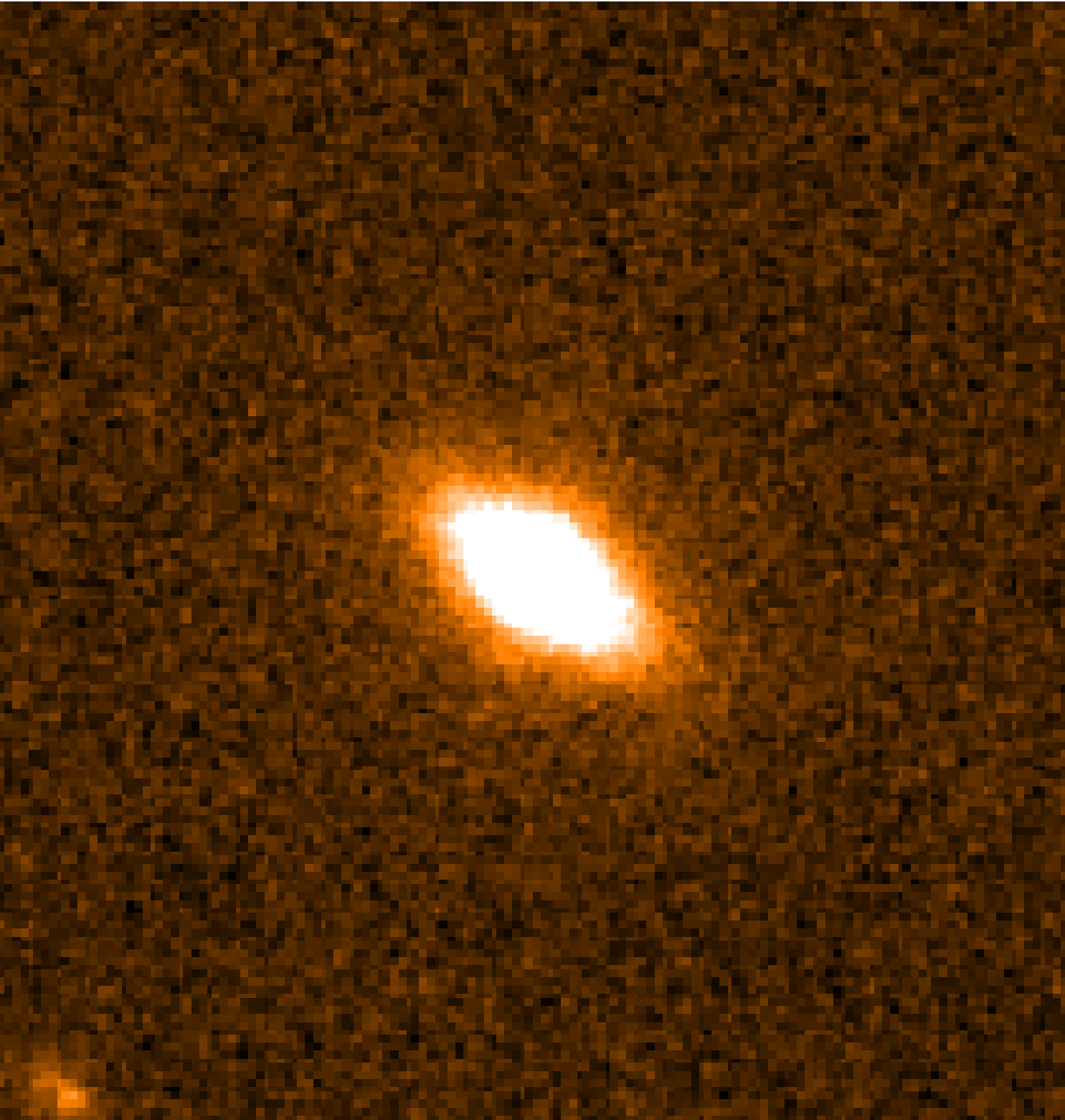}}
\quad
\subfigure
{\includegraphics[width=0.20\textwidth]{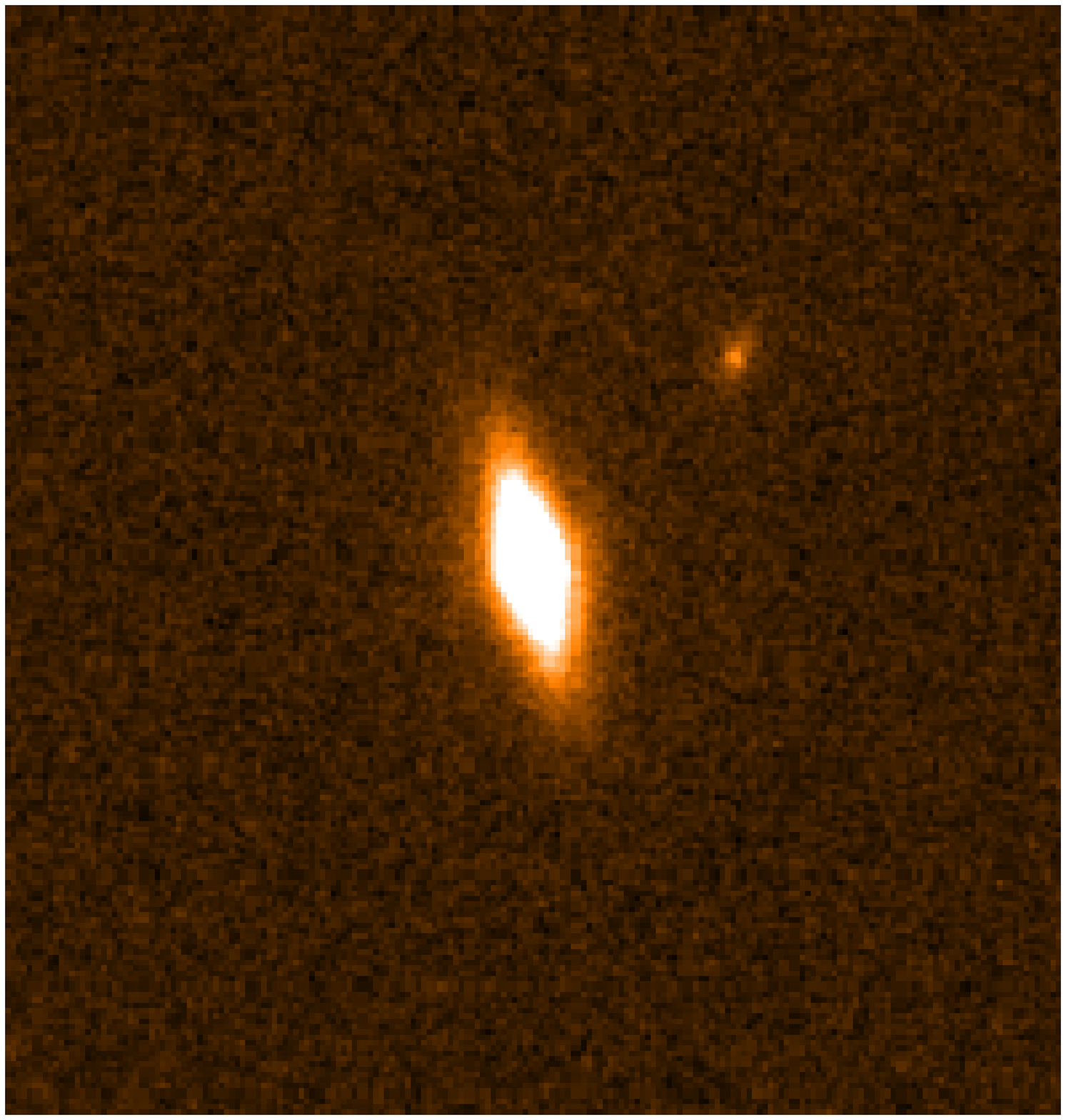}}
\quad
\subfigure
{\includegraphics[width=0.20\textwidth]{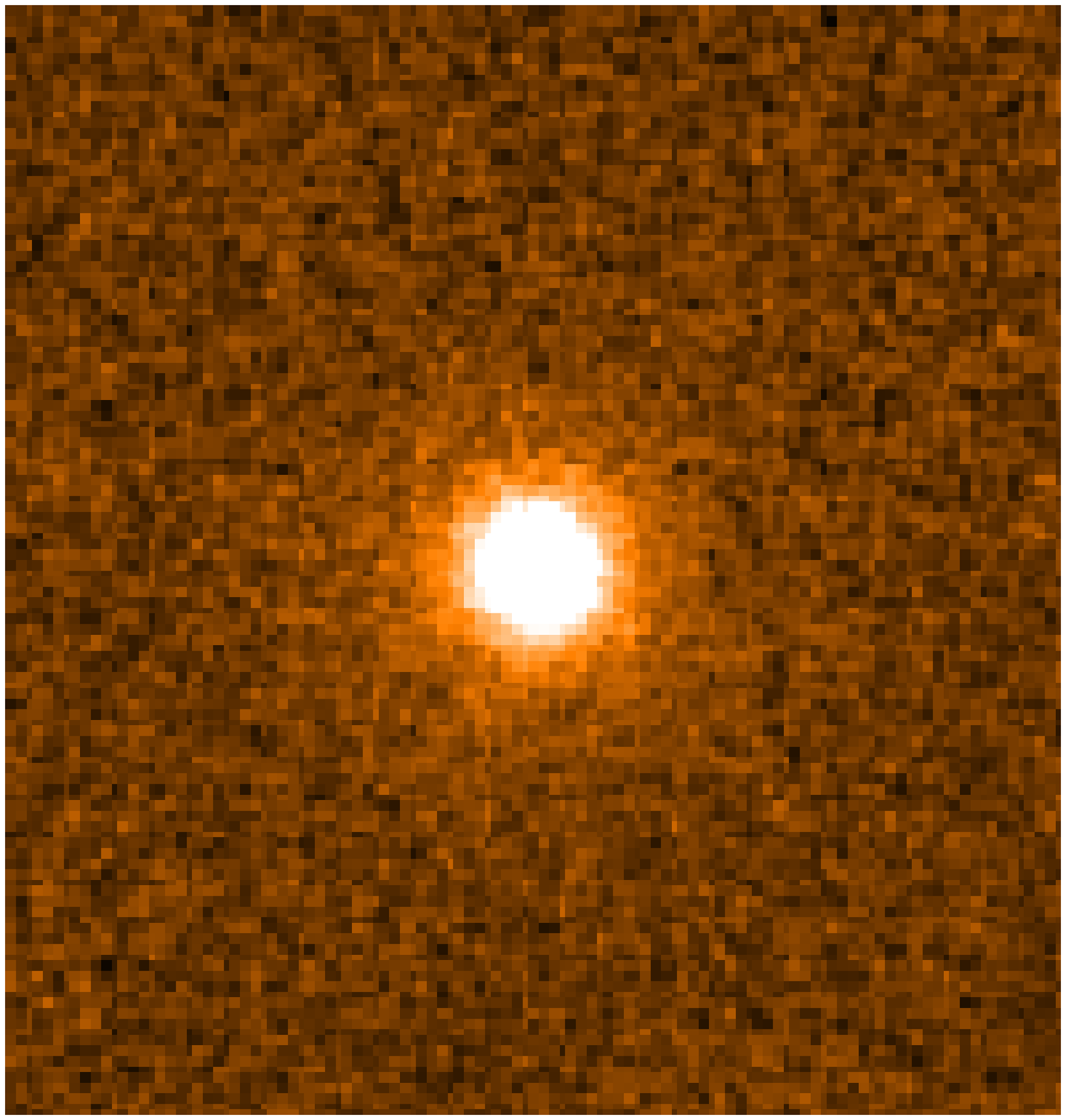}}
\\
\subfigure
{\includegraphics[width=0.20\textwidth]{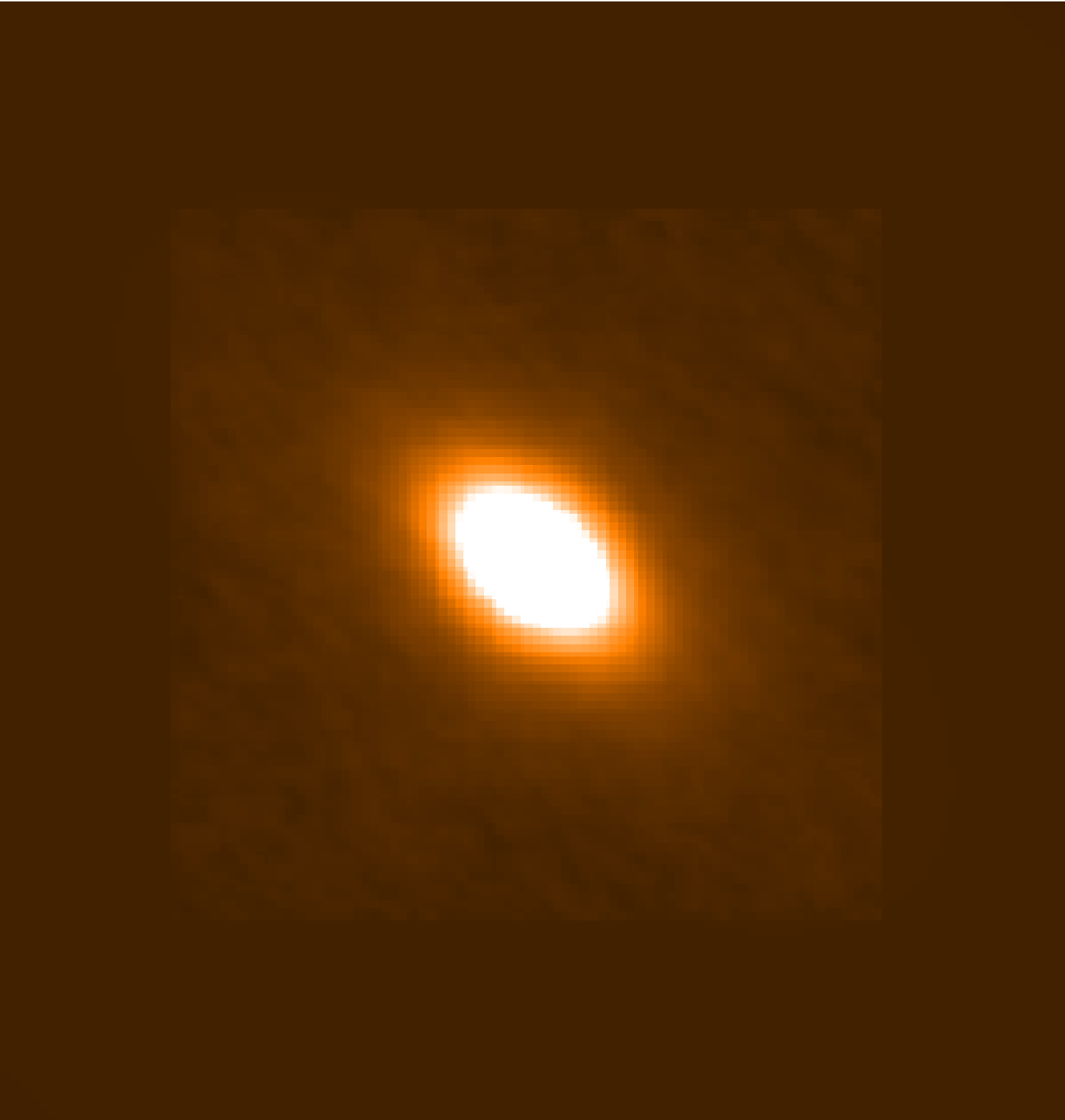}}
\quad
\subfigure
{\includegraphics[width=0.20\textwidth]{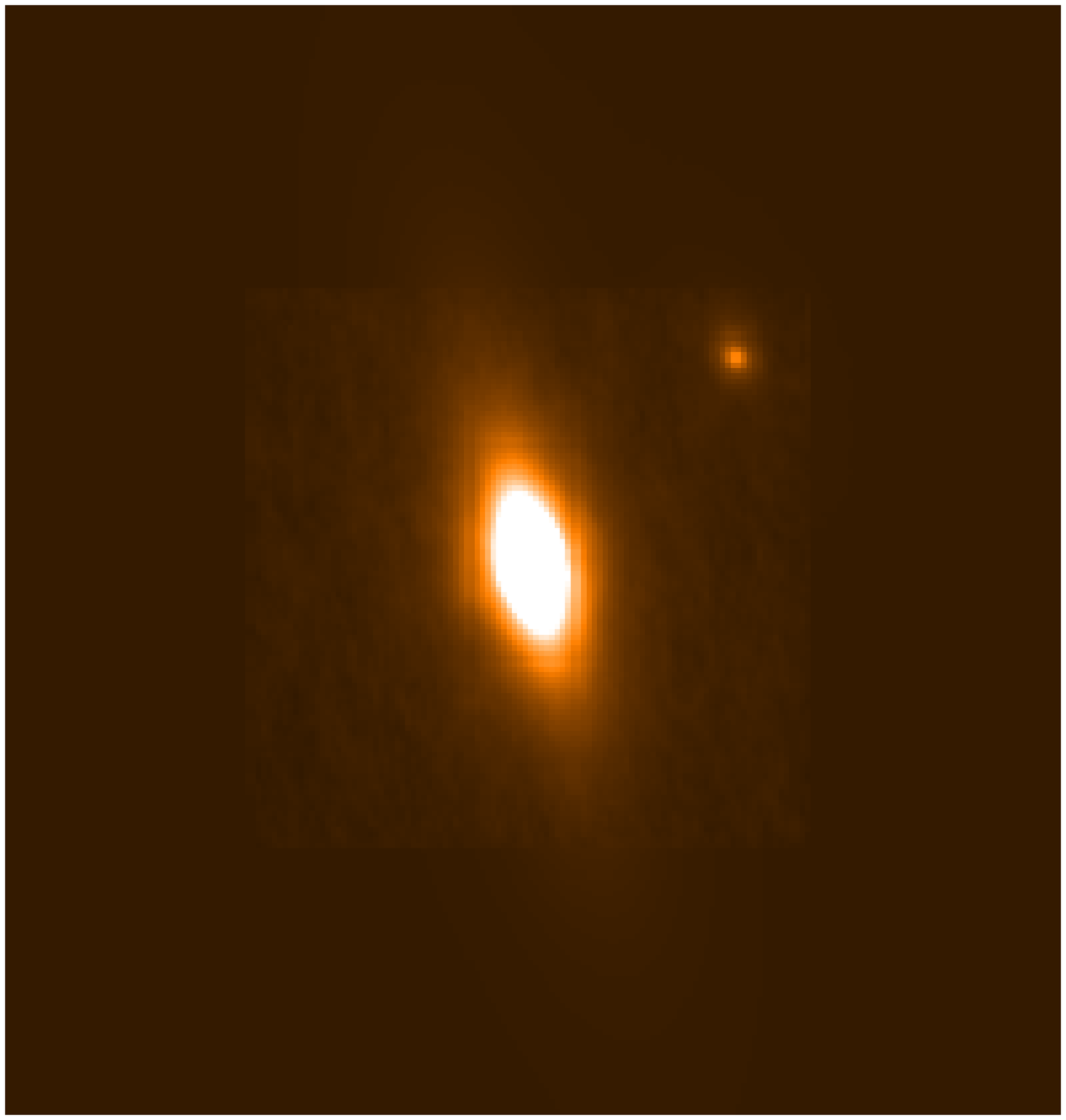}}
\quad
\subfigure
{\includegraphics[width=0.20\textwidth]{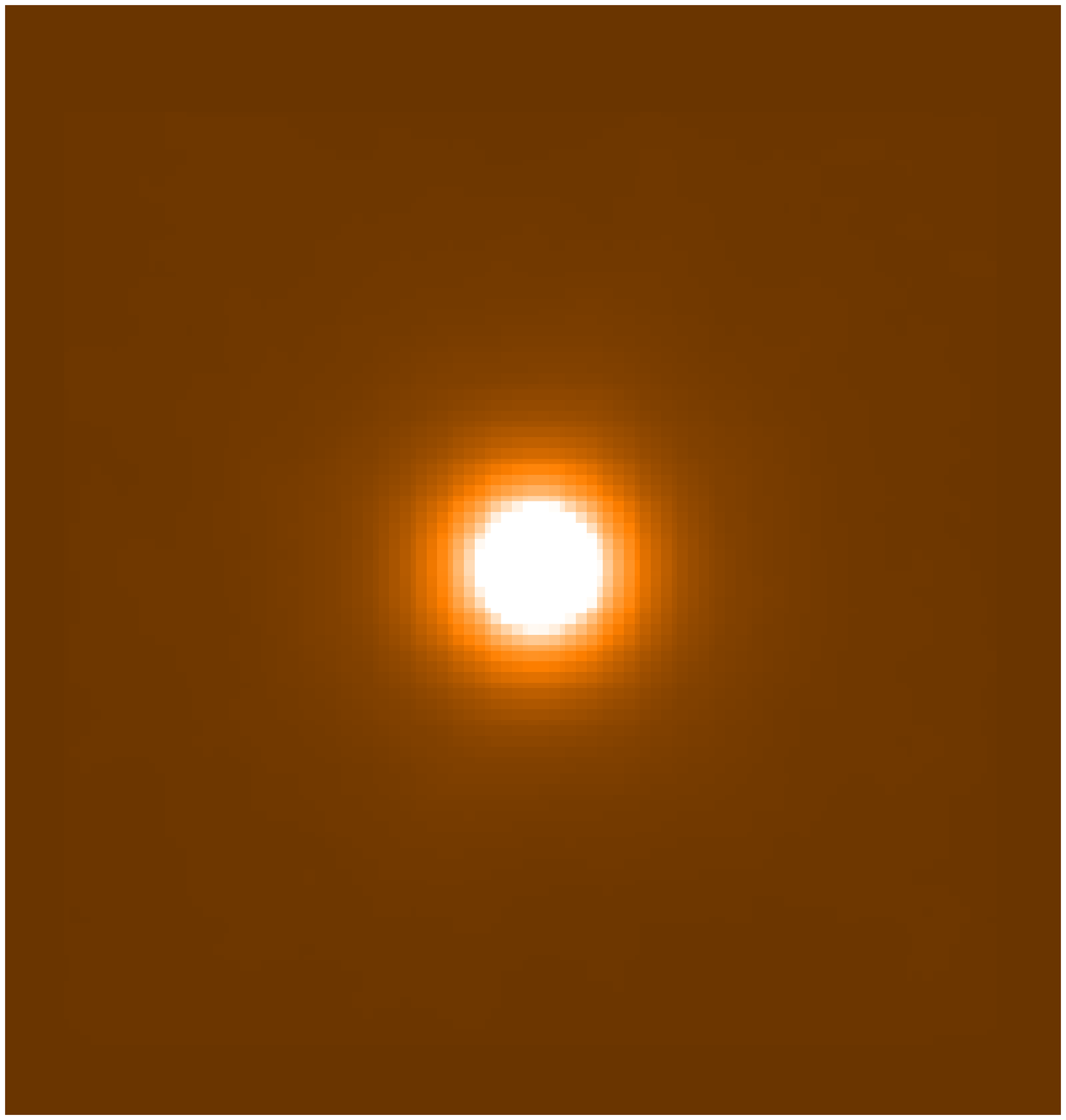}}
\\
\subfigure
{\includegraphics[width=0.20\textwidth]{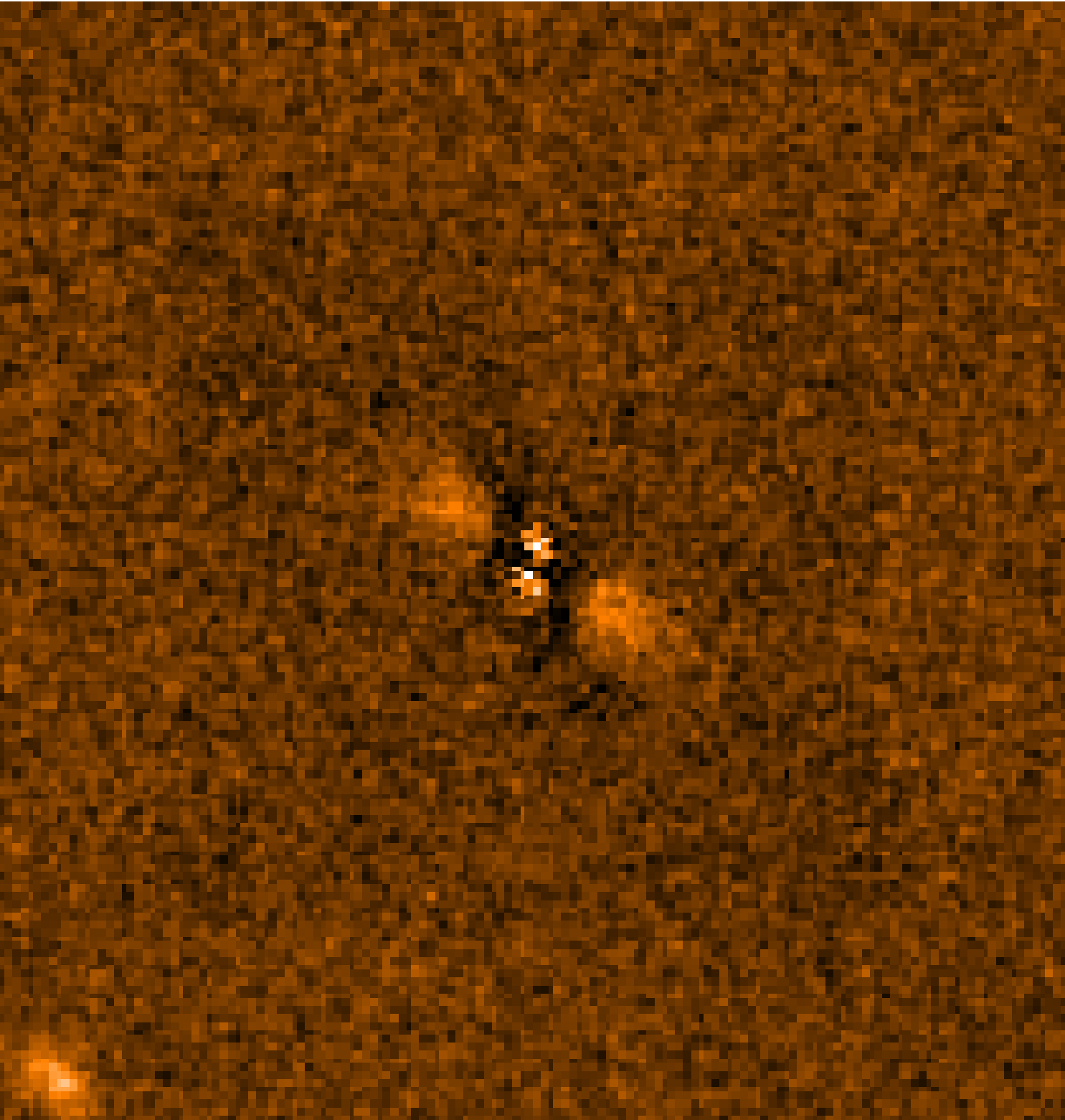}}
\quad
\subfigure
{\includegraphics[width=0.20\textwidth]{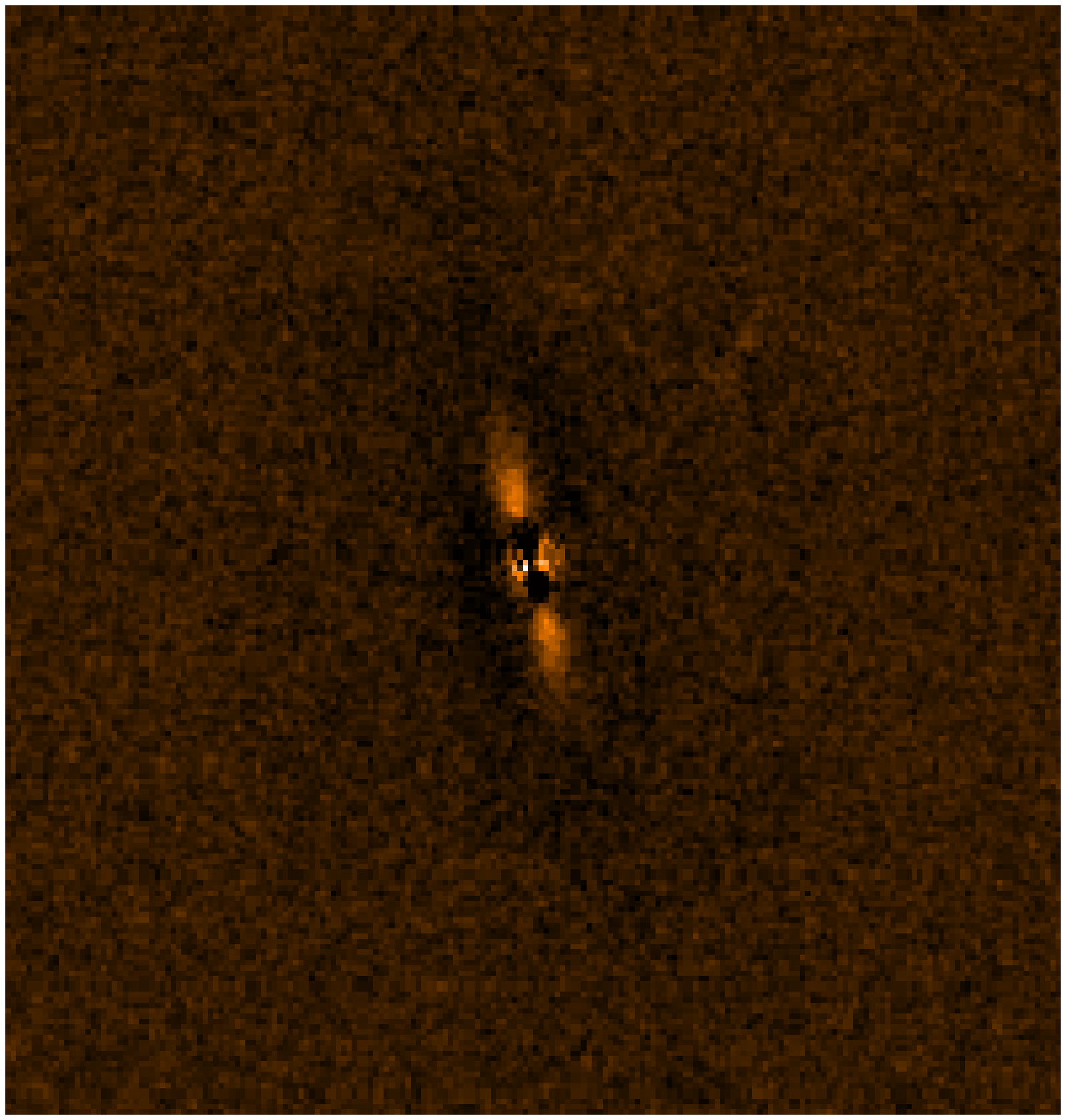}}
\quad
\subfigure
{\includegraphics[width=0.20\textwidth]{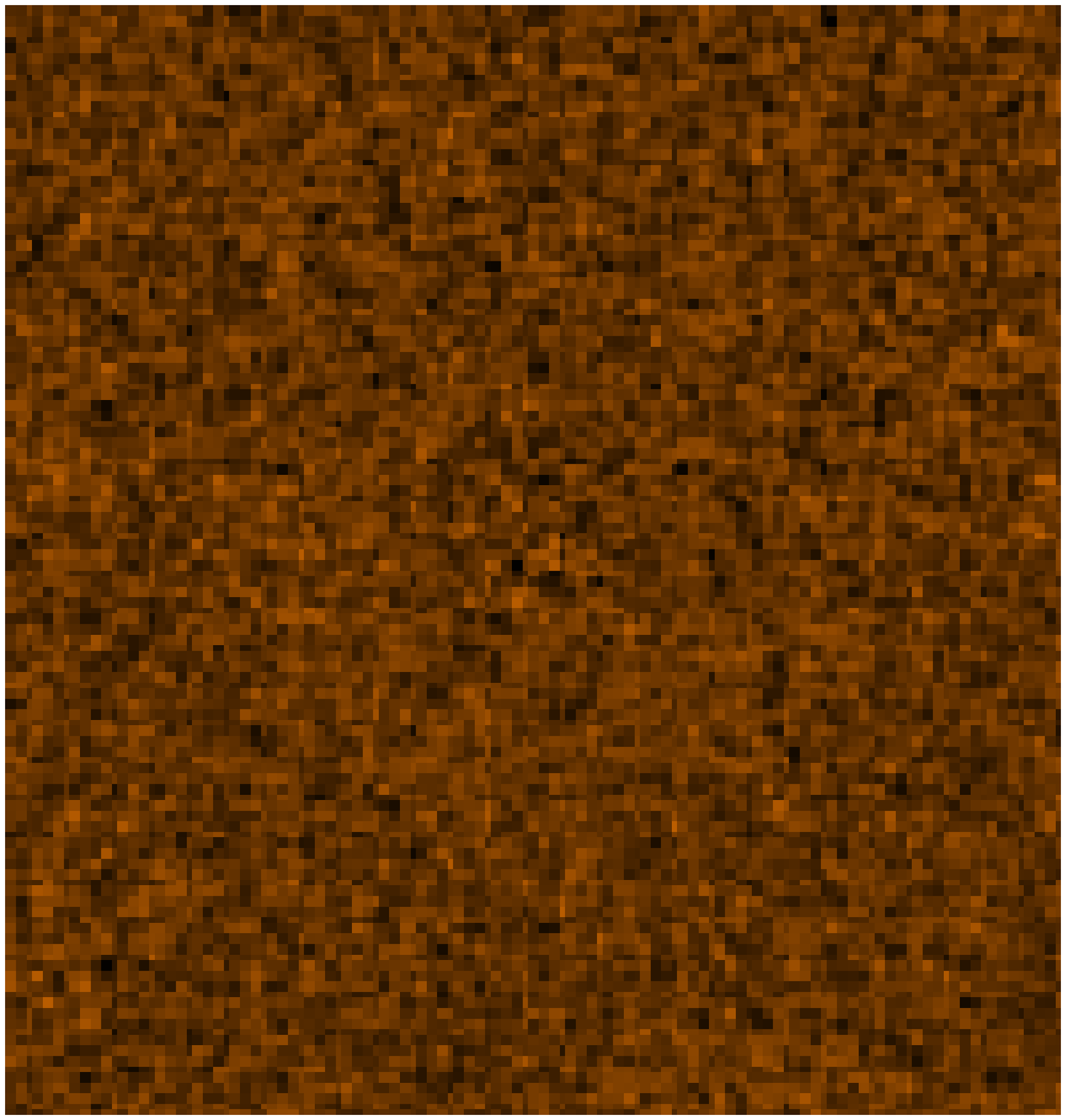}}
\\
\caption{\texttt{GALFIT} input and output for three galaxies of the magnitude limited sample. In the upper panels are the images of the galaxies in the F850LP filter, in the middle panels the best fitting S\'{e}rsic model profiles, and in the lower panels the residual images obtaining by subtracting the model from the image. The first two columns illustrate examples of late morphology galaxies with a best fitting S\'{e}rsic index $n>2.5$, while the last column displays an elliptical galaxy with $n<2.5$.}
\label{imm_galfit_bad_n}
\end{figure*}

Finally, we classified as ETGs those galaxies with a regular shape, no signs of disc in the F850LP images, and no irregular or structured residuals resulting from the fitting (e.g. Fig. \ref{imm_galfit}, bottom row). Galaxies having a clear sign of structure in the residual maps were categorised as late-type galaxies. We note that the limiting surface brightness of the ACS images in the F850LP filter ($\sim27.5$ mag/arcsec$^{-2}$) is at least 2.5 mag fainter than the faintest effective surface brightness ($\mu_e\sim25$ mag/arcsec$^{-2}$) that we measured for our ETGs. Hence, the depth of the images gives us an accurate sampling of the profile of the galaxies well beyond 2 $r_e$ for all the galaxies.

We note that in our morphological classification no selection criterion based on the S\'{e}rsic index has been adopted in the ETGs or in the LTGs identification. This choice was made after a careful analysis of the residual maps. Indeed, for a significant number of cases the fitting results, despite their goodness, provided a high value of $n$ ($n>$2.5 generally related to an elliptical profile) for clearly disc-shaped galaxies and conversely a low value of $n$ ($n<$2.5) for pure spheroidal galaxies. 
Some examples of clear misclassification that would have been made if we had selected ETGs by making a blind cut with $n>2.5$ are shown in the first two columns of Fig. \ref{imm_galfit_bad_n}. In the last column of the same figure an example of a pure elliptical that would have been excluded from the ETGs sample for a low S\'{e}rsic index ($n<$2.5) is shown.
In each column the upper panel represents the F850LP image of the galaxy, the middle panel shows the best-fitting S\'{e}rsic model profile convoluted with the PSF, and the lower panel the residual image. The presence of symmetric low surface brightness arms in the residual maps is clearly visible in the first two columns, indicating the spiral morphology of the galaxies despite the high value of the S\'{e}rsic index. In the left panel the fitting result provides values of $n=3.3$ and $\chi_{\nu}^2=1.17$, while the central galaxy has $n=4.5$ and a bad $\chi_{\nu}^2=1.80$; in this last case if we had not analysed the residual or the chi squared we would have arrived at the wrong the morphological classification. Thus, the S\'{e}rsic based criterion would not have allowed the separation between early- and late-type population. The choice to exclude this selection criterion is also in agreement with \citet{mei12}. The authors, analysing a sample of morphologically classified ETGs in clusters and groups, found that if they classify ETGs using $n>$2.5 as classification method then the ETG sample would be composed of $\sim 40\%$ of late-type galaxies.

From the morphological analysis we classified 247 out of 1302 objects as early-type galaxies in a redshift range $0.1\lesssim$z$\lesssim2.5$. We classified the remaining galaxies as LTGs.

\subsection{Selection of passive galaxies}
\label{sec:passive}
We defined passive galaxies as those galaxies with sSFR$\leq$10$^{-11}$ yr$^{-1}$. This arbitrary value allows us to compare our results with previous works, most of which use the same criterion to select passive galaxies \citep[e.g.][]{cassata11}. The sSFR is the ratio between the galaxy star formation rate (SFR) and its stellar mass (M$_*$), both derived by best fitting the observed spectral energy distribution (SED) with stellar population synthesis models.
In particular, to measure these parameters we fitted the observed SED of the 1302 galaxies over 15 photometric bands (from 0.35 $\mu m$ to 24 $\mu m$) with the stellar population synthesis models of Bruzual \& Charlot \citep[][hereafter BC03]{bruzual03} at fixed solar metallicity.
The fits were performed with the software \texttt{HYPERZMASS}, a modified version of the photometric redshift code \texttt{HYPERZ} \citep{bolzonella00}, which finds the best-fitting template with a standard $\chi^2$ normalisation at fixed known redshifts. For the 24 \% of the galaxies ($\sim 310$) without spectroscopic redshift we adopted the photometric $z$ from \citet{santini09} whose accuracy is about $\Delta z/(1+z)\simeq0.045$.
Eight star formation histories (SFHs) described by an exponentially declining star formation rate (SFR$\propto$exp$^{-t/\tau}$) with e-folding time $\tau$=[0.1, 0.3, 0.4, 0.6, 1, 2, 5, 15] Gyr and 221 ages in the range t=[0.001, 20] Gyr were used, where the age of the template at each redshift was not allowed to exceed the age of the Universe at that redshift. 
The Calzetti law \citep{calzetti00} was applied to describe the dust extinction and the extinction parameter $A_{V}$ was fitted in the range $0\leq A_{V}\leq$1.8 mag imposing the prior $A_{V}>0.6$ only if $t/\tau<$4, i.e. allowing significant extinction only for galaxies with high SFR.

\begin{figure}
{\includegraphics[width=8.5cm]{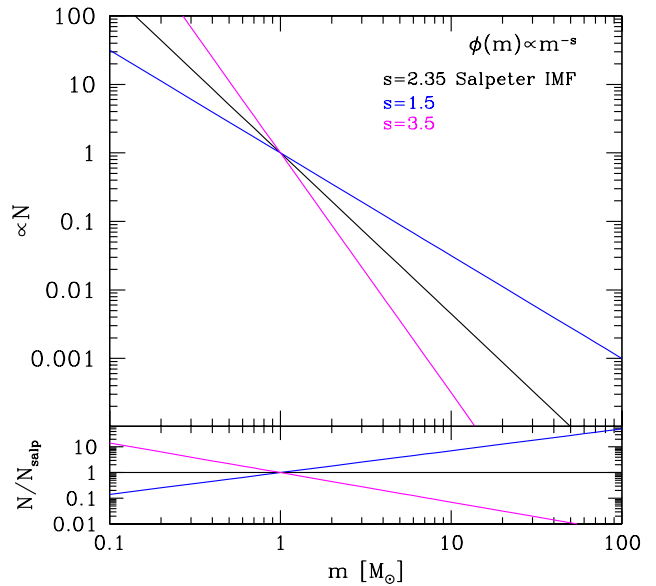}}
\caption{Stellar initial mass functions with a different proportion of low-to-high mass stars ($m$) described by a power-law IMF $\phi(m)\propto m^{-s}$ with slopes s=1.5, 2.35 (Salpeter IMF), and 3.5. The steeper the slope of the function is, the higher the contribution of low-mass stars.}
\label{fig:imf}
\end{figure}

For each of the 1302 galaxies of the sample, the fitting process was repeated assuming seven different IMF: the classical Salpeter \citep{salpeter55} and Chabrier IMFs \citep{chabrier03}, and five other power-law IMFs ($\phi(m)\propto m^{-s}$) with slope \textit{s}=1.5, 2.0, 2.5, 3.0, 3.5. The grid of stellar population models based on the last five IMFs were kindly provided by Stephan Charlot, with lower and upper mass ($m$) cutoffs of 0.1 and 100 M$_{\odot}$. In Fig. \ref{fig:imf} the three IMFs with slope s=1.5, 3.5, 2.35 (Salpeter IMF) are shown to highlight the different percentage of low-mass stars among them.
From the SED fitting analysis we derived for each galaxy its stellar mass M$_*$, the mean age of its stellar population, and its star formation rate. The stellar mass we adopt is the mass associated with the best fitting template considering only the mass contained within stars, including stellar remnants (white dwarf, neutron stars, etc.).
The best fitting templates of 90 \% of our magnitude limited sample, obtained with a Salpeter IMF, are defined by SFHs with $\tau \leq 1$ Gyr, ages t$<3.5$ Gyr, and stellar masses in the range $10^9<$M$_*<2\times 10^{11}$ M$_{\odot}$.

\begin{figure}
\includegraphics[width=8.5cm]{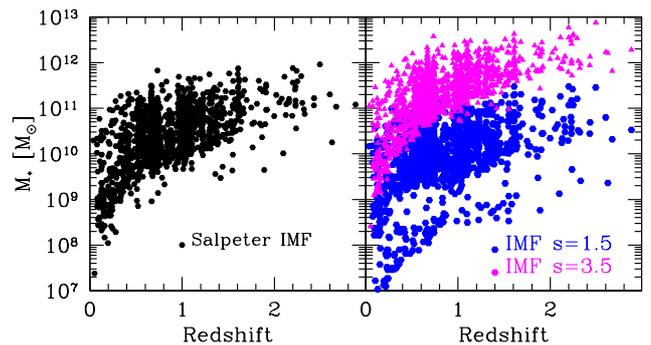}
\caption{Stellar mass as a function of redshift obtained by assuming a Salpeter initial mass function in the spectrophotometric models (\textit{left panel}) and two Salpeter-like IMFs ($\phi(m)\propto m^{-s}$) with a lower (s=1.5) and higher (s=3.5) abundance of low-to-high mass stars (\textit{right panel}).}
\label{fig:massa_limite}
\end{figure}

The stellar mass of the galaxies as a function of their redshift is shown in Fig. \ref{fig:massa_limite}. The left panel of Fig. \ref{fig:massa_limite} shows the stellar mass derived assuming a Salpeter IMF in the SED fitting, while the right panel of Fig. \ref{fig:massa_limite} refers to stellar mass obtained using a power law IMF (Salpeter-like) with slopes s=1.5 and s=3.5. 
The dependence of the stellar mass on the IMF adopted in the SED fitting is well known. Indeed the stellar mass of a galaxy is mainly determined by the long-lived low-mass stars ($<1$ M$_{\odot}$). Hence it follows that any variation of the relative abundance of low-to-high mass stars drastically modifies the basic properties of a galaxy.

Applying the sSFR selection (sSFR$\leq$10$^{-11}$ yr$^{-1}$) we extracted a sample of 281 passive galaxies, assuming in the models a Salpeter IMF.
From now on, unless otherwise specified, we refer to passive galaxies derived using a Salpeter IMF for the comparison with the morphologically selected sample. In Sect. \ref{sec:imf_passive} we will discuss how a passive galaxy selection depends on the different IMFs we adopted in the spectrophotometric models.

\section{Comparison between morphological and passive samples}
The selection of passive galaxies led to a sample of 281 galaxies while the morphological selection produced a sample of 247 ETGs. Actually, the mismatch between these numbers is not the most important difference between the two selection criteria. 
Indeed, of the 247 visually classified ETGs, 182 are passive galaxies while 65 ETGs do not satisfy the passivity criterion. Instead out of the 281 passive galaxies 99 have a disc-like/irregular morphology. 
This means that $\sim$35 \% of passive galaxies are not ETGs and that the selection criterion based on the passivity of the galaxies misses 26 \% of spheroids. Thus the two different criteria do not select the same sources, i.e. the sample of ETGs does not coincide with that of passive galaxies and the passive sample does not include all the pure ETGs contained in the original magnitude limited sample. 
\begin{figure*}
\includegraphics{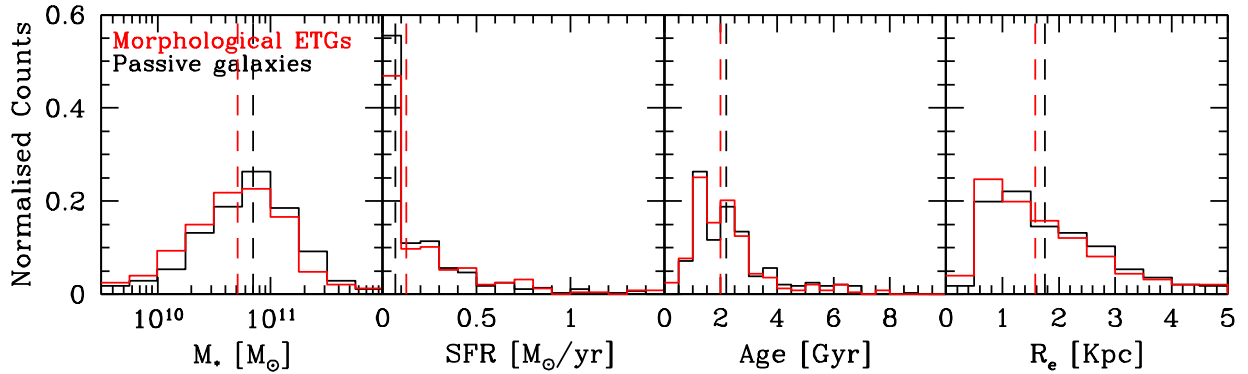}
\caption{From left to right: stellar mass, star formation rate, age of stellar populations, and effective radius distributions of passive (black) and morphological early-type galaxies (red). Vertical lines are the median values.}
\label{fig:pas_vs_morfo}
\end{figure*}
On the basis of these differences, we turn our attention to understanding if the two samples share the same physical properties. The distributions of physical and structural parameters are shown in Fig. \ref{fig:pas_vs_morfo}, where red and black represent ETGs and passive galaxies. 
We found that from the analysis of stellar masses, star formation rates, and mean ages of the stellar populations no significant differences are detected. Similarly, the distributions of the effective radii are almost indistinguishable for the two samples. This may be due to the 182 galaxies in common between the two samples; therefore, the possible differences resulting from the unshared objects are diluted in the global distributions.

For this reason we analysed the galaxies not in common between the passive and the ETGs samples (i.e. 65 non-passive ETGs and 99 passive LTGs), searching for any peculiarities that make them different from the 182 galaxies present in both samples.

With regard to the 65 pure-ETGs that fall outside the passivity criterion, we verified that $\sim$70 \% of them have a sSFR$<5\times10^{-11} \ \rm yr^{-1}$, which is slightly higher than the chosen limit. Hence these galaxies are basically equal to the passive ETGs, but because it was necessary to apply a sharp cut to the sSFR some of them are excluded from the selection.

For the 99 passive LTGs, we compared their stellar mass, age of their stellar populations, and their effective radius with those of passive ETGs. We found that the stellar masses of passive LTGs do not show measurable differences compared to the stellar masses of the passive ETGs. This is not surprising if we consider the galaxy luminosity function \citep[e.g.][]{salimbeni08}, where the ETGs at higher mass are numerically more abundant, but the mass distribution is spread over the same range between LTGs and ETGs, thus the stellar mass distribution of the passive galaxies has no dependence on the galaxy morphology.

In addition the distributions of the effective radii of discs and spheroids are very similar; for the passive late-type population, 50\% of the emitted total light is concentrated in a median radius $r_e\sim 1.6$ kpc compared to the median value $r_e\sim 1.8$ kpc of passive ETGs. To determine whether this analogy of the effective radii was expected, we calculated the effective radii of two galaxies assumed to have the same stellar mass and surface brightness profiles described by a pure disc profile ($n=1$) in one case, and by a de Vaucouleurs profile ($n=4$) in the other. We found that the  effective radii differ by a factor of 1.26, in agreement with the observed $r_e$ differences.
This means that from an analytical point of view the effective radius of a S\'{e}rsic profile generally associated with a disc ($n = 1$) does not show substantial differences compared to the $r_e$ of a profile generally associated with spheroidal object.

The two samples have almost the same age distributions, with a median age of 2 Gyr and 2.5 Gyr for ETGs and LTGs,  respectively. Again, this is not surprising because all the passive galaxies, both late- and early-type, were selected according to the same criterion of passivity, hence their SED is fitted by similar spectrophotometric templates that provide comparable age measures.

\begin{figure}
{\includegraphics[width=8.5cm]{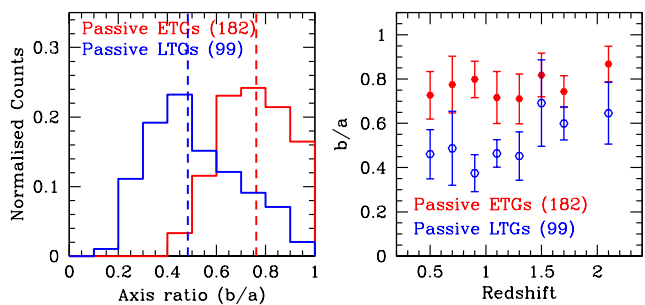}}
\caption{\textit{Left)} Axis ratio distribution of passive ETGs (red) and passive LTGs (blue). \textit{Right)} Median axis ratio evolution as a function of redshift for passive ETGs (red) and passive LTGs (blue).}
\label{fig:ba}
\end{figure}

The only differences between passive ETGs and passive LTGs have been found in their axis ratios, a parameter directly related to the shape and dynamic of the galaxy. From the left panel of Fig. \ref{fig:ba} it is clear that the two populations have different axis ratio values, where the visually classified ETGs typically have $b/a=0.76$, while the galaxies classified as discs have a median $b/a=0.48$. 
The result is consistent with previous studies at low redshift (Padilla \& Strauss 2008) and at high redshift \citep{buitrago13}; in particular in the latter case, a visually classified sample of massive galaxies at redshift $0<z<3$ has been analysed and axis ratio values are in agreement with our median results for both early- and late-type populations. Van der Wel et al. (\citeyear{vanderwel11}) analysed a sample of passive galaxies and found that a fraction of quiescent galaxies had small axis ratio values; they claimed that these measurements were probably tracers of a disc population among the quiescent galaxies, but they did not obtain the morphological information for these objects. Our analysis confirms this clue and also provides the percentage of disk-like objects in a sample of passive galaxies.
In the left panel of Fig. \ref{fig:ba} the presence of an overlapping region between the axis ratios of ETGs and LTGs can also be seen, with a fraction of discs populating the tail at greater axis ratios. Since at high redshift the profile fitting may be biased by the low signal-to-noise ratio of the images \citep{mancini10} because the surface brightness scales as $(1+z)^4$, we investigated if the higher $b/a$ values of LTGs were due to a systematic loss of the low surface brightness structures at increasing redshift.
The result shown in the right panel of Fig. \ref{fig:ba} suggest that this may affect the measurement of the axis ratios of passive LTGs that tend to be greater at higher redshift. However, we cannot exclude that the observed trend is due to a physical evolution of the galaxy shapes.

The analysis of the axis ratios provides additional and independent information on the morphology of the galaxies and confirms that the passive sample is composed of morphologically different galaxies, which cannot be handled as homogeneous population of objects.

\section{Dependence of the passivity of galaxies on the IMF}
\label{sec:imf_passive}
The definition of passive galaxies rests on a selection criterion based on the sSFR, a model-dependent quantity, and on an arbitrary threshold value of this quantity (e.g. sSFR$\leq10^{-11} \ \rm yr^{-1}$). The computation of the sSFR results from the ratio between the star formation rate and the stellar mass, both of which are derived from the fitting of the observed SED with stellar population synthesis models. Here, we investigate how the samples of passive galaxies and their morphological mix change, at fixed sSFR thresholds and for different IMFs adopted in the models. In literature the same threshold of sSFR is often adopted with different IMFs, not considering that the variation of the relative abundance of low-to-high mass stars changes the stellar mass and the SFR with which the sSFR is estimated. Thus, we wonder how the sSFR of a galaxy depends on the assumed IMF and how the same definition of passivity (i.e. sSFR$\leq10^{-11} \ \rm yr^{-1}$) affects the selection of samples of passive galaxies.

To this end, from the original magnitude limited sample of 1302 galaxies we selected five samples of passive galaxies, one for each of the five power-law slopes (s=1.5, 2.0, 2.5, 3.0, 3.5) of the considered IMF (see Sec. \ref{sec:passive}), according to the selection criterion sSFR$\leq 10^{-11} \ \rm yr^{-1}$.

The results of the selection show that the higher the abundance of low-mass stars in the IMF, the lower the number of objects classified as passive galaxies, from 146 objects for the bottom-heavy IMF (s=3.5) to 403 for the top-heavy IMF (s=1.5).
This trend could be due to a balance of the best fitting parameters: to fit a given observed SED, the intrinsically redder colour of the models with steeper IMF must be counterbalanced by younger ages and/or longer SFHs. As a consequence, higher value of sSFR than the above selection criterion will be assigned to the galaxy.

Since the galaxies collected in the various samples are different, we analysed how the morphological composition of these samples (i.e. the number of ETGs and LTGs) changes as a function of the IMF.
We found that the passive galaxies are not all spheroidal galaxies and that irrespective of the assumed IMF a remarkable fraction of passive galaxies have a late-type morphology.
In particular, for increasing abundance of low-mass stars, the selected ETGs move from $\sim85\%$ (i.e. 209/247 for s=1.5) to $\sim 41.7\%$ (i.e. 103/247 for s=3.5) when compared to the 247 ETGs of the morphological sample. However, although the samples with steeper IMF select a smaller number of ETGs, they have a lower contamination of discs; the fraction of LTGs passes from 48\% in the sample with IMF$_{s=1.5}$ to a 29\% regarding the sample obtained with IMF$_{s=3.5}$. Hence, keeping the sSFR threshold fixed and varying the IMF, the selected samples of passive galaxies considerably differ both in the number of selected galaxies and in the morphological composition.
 \begin{table}
 \label{tab:1}
 \centering
 \begin{tabular}{llccc}\hline
IMF Type & sSFR threshold & Passive & ETGs & LTGs\\ 
 & [$yr^{-1}$] & Galaxies & fraction & fraction\\ \hline\hline
Salpeter & $\leq1.0\times10^{-11}$ & 281 & 182 & 99\\
Chabrier & $\leq1.2\times10^{-11}$ & 295 & 184 & 111\\
s=1.5 & $\leq5.7\times10^{-12}$ & 319 & 181  & 138\\
s=2.0 & $\leq7.7\times10^{-12}$ & 313 & 193  & 120\\
s=2.5 & $\leq1.4\times10^{-11}$ & 283 & 182  & 101\\
s=3.0 & $\leq3.1\times10^{-11}$ & 280 & 179  & 101\\
s=3.5 & $\leq7.1\times10^{-11}$ & 286 & 182  & 104\\ \hline
 \end{tabular} 
 \caption{Samples of passive galaxies with relative morphological compositions, selected using different sSFR thresholds, according to the different IMFs adopted in the population synthesis models.}
 \end{table} 

For this reason, a great deal of attention should be paid to the choice of the sSFR threshold in order to select and compare samples of passive galaxies when different IMFs are used. In particular, to be consistent, the sSFR threshold should be rescaled according to the IMF used. We performed this exercise by adopting as reference the sSFR derived assuming a Salpeter IMF and we redefined the sSFR threshold for the other IMFs. In particular, for each IMF  we derived a scale factor as the median value of the ratio $sSFR_{IMF}/sSFR_{Salp}$ obtained for each of the 1302 galaxy. The results are shown in Table \ref{tab:1}.
The new sSFR thresholds are sSFR$_{s=1.5}\leq5.7\times10^{-12} \ \rm yr^{-1}$, sSFR$_{s=2.0}\leq7.7\times10^{-12} \ \rm yr^{-1}$, sSFR$_{s=2.5}\leq1.4\times10^{-11} \ \rm yr^{-1}$, sSFR$_{s=3.0}\leq3.1\times10^{-11} \ \rm yr^{-1}$, sSFR$_{s=3.5}\leq7.1\times10^{-11} \ \rm yr^{-1}$, and sSFR$_{Chabrier}\leq1.2\times10^{-11} \ \rm yr^{-1}$. With these new cuts in sSFR we repeated the selection of the passive samples and we found that the number of passive galaxies varies from a minimum of 280 with an IMF$_{3.0}$ to a maximum of 319 with an IMF$_{1.5}$ (Col. 3 in Table \ref{tab:1}). The analysis of the number of ETGs and LTGs included in each of the five new samples shows that none of these selections of passive galaxies allow us to collect all the ETGs regardless of the IMF assumed in the models. Moreover, all the samples contain a considerable fraction ($\sim$36-43 \%) of late-type populations. In conclusion, a significant number of LTGs is included in the passive population, whatever the IMF assumed in the models, and this should be accounted for when studying the evolution of the properties of the ETGs population.

\section{Following the ETG and LTG evolutions through their main properties}
\begin{figure*}
\includegraphics[width=17 truecm]{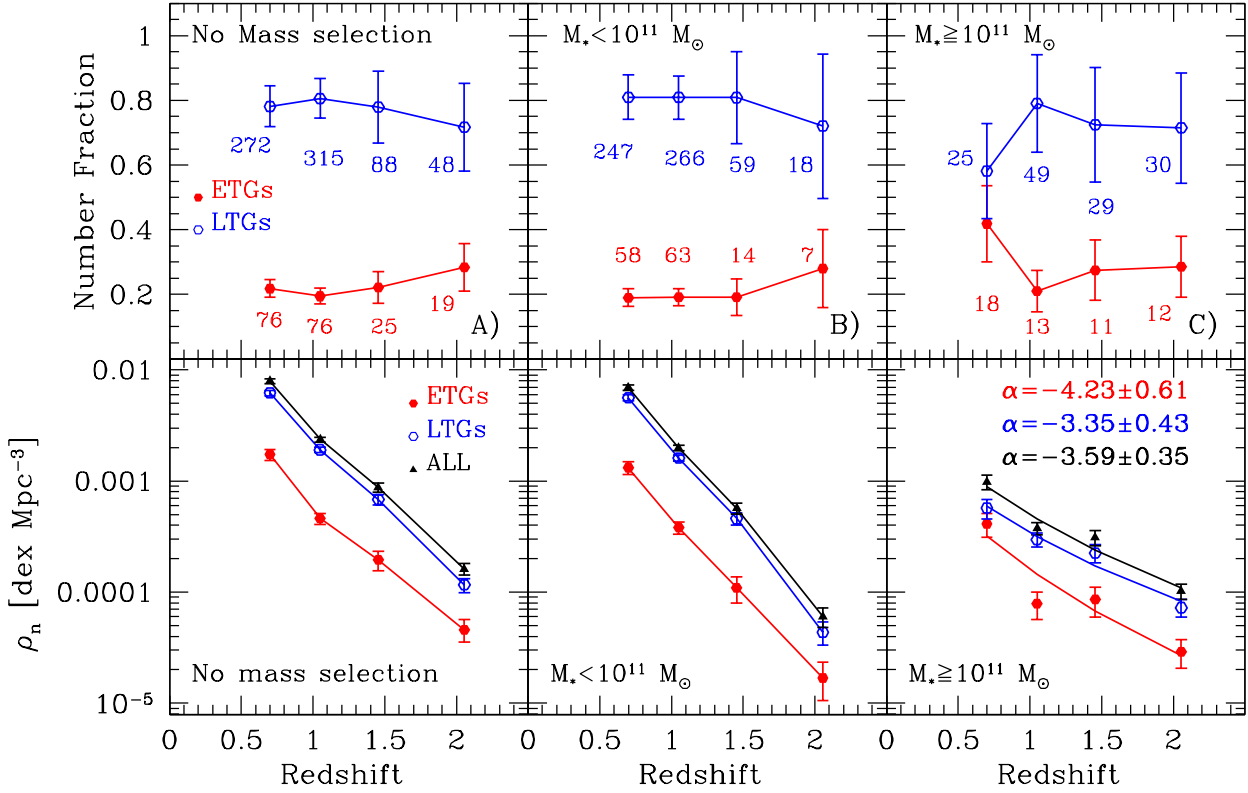}
\caption{First line. Evolution of the morphological fraction of ETGs (red) and LTGs (blue) for the entire morphological sample A), for the low-intermediate mass sample B) and for the massive sample C). The number of galaxies in each redshift bin is given. Second line. Cosmological evolution of the ETG and LTG number density as a function of redshift in the same mass ranges as the first line panels. Black lines represent the number density for the entire sample.}
\label{fig:fra_num_den}
\end{figure*}
We have shown that the selection of passive galaxies strongly depends on the IMF assumed in the models, is affected by LTGs contamination, and misses a significant fraction of the population of ETGs. 
Since morphology and passivity are not equivalent we cannot use samples of passive galaxies to study the evolution with redshift of a particular morphological class. 

In this section we study the evolution with redshift of the number of galaxies and of the stellar mass content of early-type galaxies to put constraints on their mass assembly. For this purpose we use the morphological sample that we obtained on the basis of the visual classification described in Sec. \ref{sec:morpho}, composed of late-type objects and strictly defined ETGs.

We considered two mass ranges in our analysis: M$_*\geq10^{11}$ M$_\odot$, for which the sample is complete over the whole redshift range (hereinafter we refer to this sample as the massive galaxies sample), and M$_*<10^{11}$ M$_\odot$, the range of low-intermediate mass where the completeness threshold of M$_*$ is different at different redshift, as shown in Fig. \ref{fig:massa_limite}. Since the small field of view of the GOODS-S survey does not allow a good sampling of the high mass galaxies for $z<0.6$ (Fig. \ref{fig:massa_limite}), to ensure the mass completeness in the sample of massive galaxies we have also applied a redshift restriction in the range $0.6\leq z \leq2.5$; this redshift cut was then extended to the low-intermediate mass galaxies. The final flux limited sample considered at $0.6\leq z \leq2.5$ is composed of 196 ETGs and 723 LTGs.

 \begin{table*}
 \label{tab:fraction}
 \centering
 \begin{tabular}{lllll}\hline
Redshift & ETGs fraction & LTGs fraction & ETGs fraction & LTGs fraction\\
 & M$_*<10^{11}$ M$_\odot$ &  M$_*<10^{11}$ M$_\odot$ & M$_*\geq10^{11}$ M$_\odot$ & M$_*\geq10^{11}$ M$_\odot$ \\ \hline\hline
$0.6\leq z\leq0.8$ & $0.19\pm0.03$ &$0.81\pm0.07$  & $0.42\pm0.12$  &$0.58\pm0.15$\\
$0.8<z\leq1.3$     & $0.19\pm0.03$ &$0.81\pm0.07$  & $0.21\pm0.06$  &$0.79\pm0.15$\\
$1.3<z\leq1.6$     & $0.19\pm0.06$ &$0.81\pm0.14$  & $0.27\pm0.09$  &$0.73\pm0.18$\\
$1.6<z\leq2.5$     & $0.28\pm0.12$ &$0.72\pm0.22$  & $0.29\pm0.09$  &$0.71\pm0.17$\\ \hline
 \end{tabular} 
 \caption{Number fraction of ETGs and LTGs in the low (M$_*<10^{11}$ M$_\odot$) and high mass (M$_*\geq10^{11}$ M$_\odot$) regimes for four redshift bins.}
 \end{table*}

In Fig. \ref{fig:fra_num_den} (top row) the fraction of the two morphological classes is shown as a function of redshift for the whole sample (panel A), the low-intermediate mass range sample (M$_*<10^{11}$ M$_\odot$, panel B) and the massive galaxies sample (M$_*\geq10^{11}$ M$_\odot$, panel C). The data are given in Table 2. The most evident result from Fig. \ref{fig:fra_num_den} A) and B) is that the fraction of ETGs and LTGs is nearly constant over the whole redshift range considered. In particular the late-type objects are the dominant population at all redshifts, being more than 70 \% of the entire population, while the spheroidal fraction remains constant at about 20-30 \%.
\begin{figure*}
\includegraphics[width=7.5 truecm]{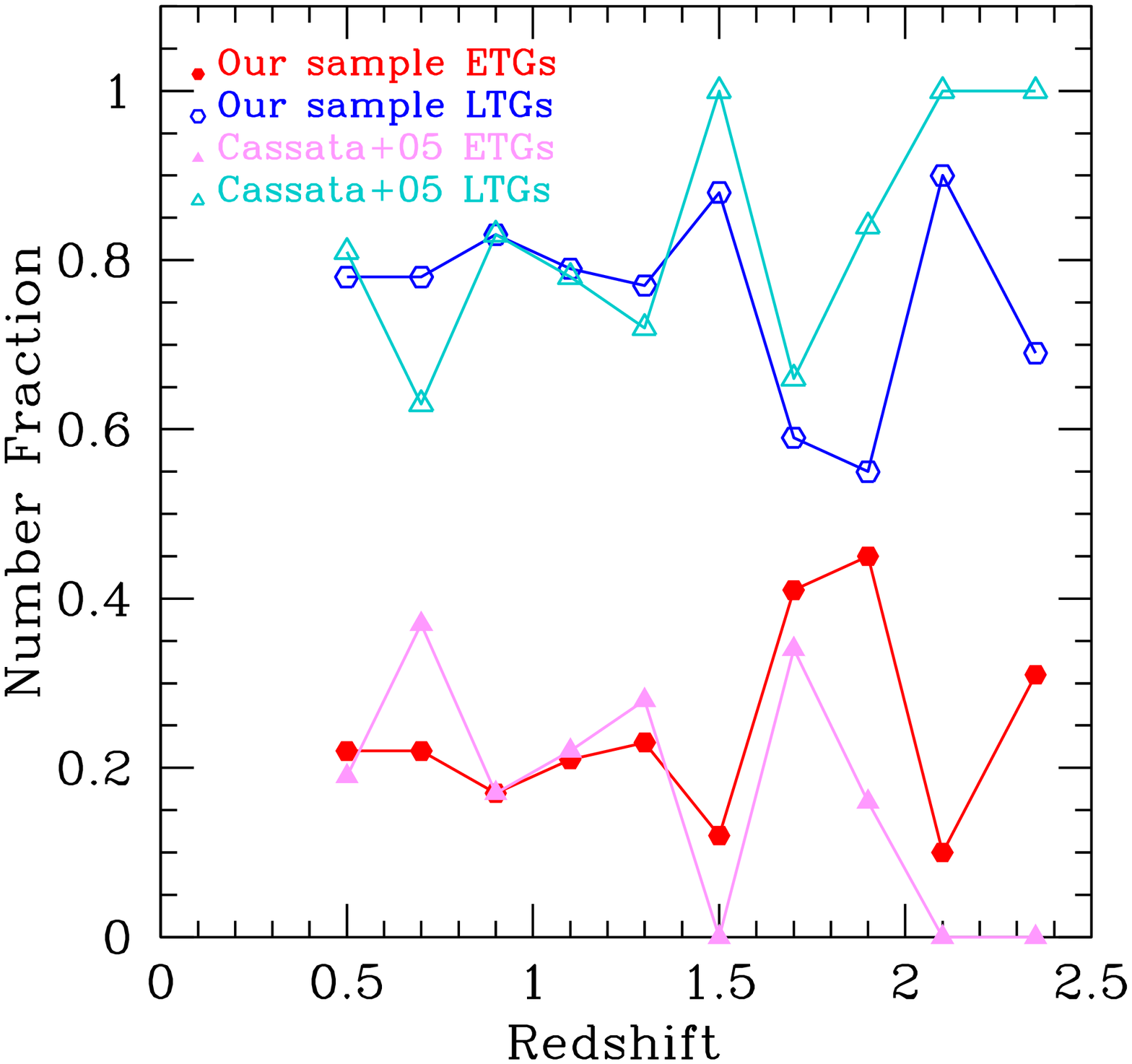}
\includegraphics[width=7.5 truecm]{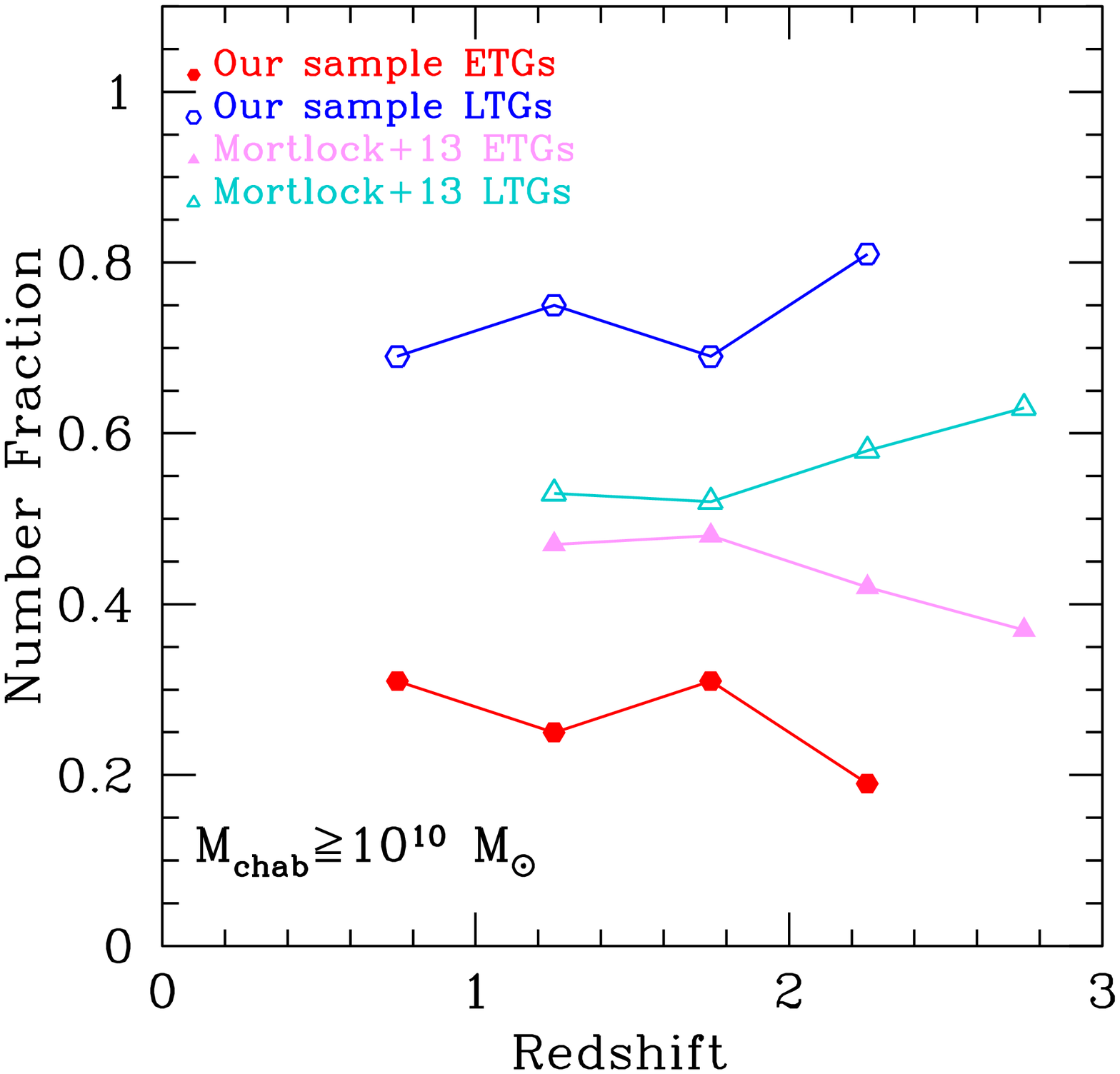}
\caption{Comparison of the morphological fractions of our morphological sample (ETGs in red, LTGs in blue) with the ones derived by \citet{cassata05} (left panel) and \citet{mortlock13} (right panel). ETGs and LTGs of the other authors are shown in pink and cyan, respectively. The redshift binning is chosen in order to best reproduce the other works intervals.}
\label{fig:confronti}
\end{figure*}
Our result is in agreement with that of \citet{cassata05}, based on a morphologically selected sample in a small portion of the GOODS-S field, as shown in the left panel of Fig. \ref{fig:confronti}. This agreement provides evidence of the reproducibility and hence of the reliability of the visual classification since the data refer to the same survey but the classification is independent.

We also compared our result with \citet{mortlock13}; the authors use a sample of galaxies at $1\leq z\leq3$ with $M_*\geq10^{10} M_{\odot}$ in the CANDELS/UDS field to investigate the formation of the Hubble sequence.
As can be seen from the right panel of our Fig. \ref{fig:confronti},  \citet{mortlock13} find that the two classes of galaxies are almost equivalently populated in disagreement with our findings. This inconsistency may be due to the band used for the visual classification performed by the authors, the H$_{160}$-band, a redder band with a lower resolution compared to the F850LP-band, that can produce the loss of structures more visible in a bluer band.

\begin{figure*}
\includegraphics[width=19 truecm]{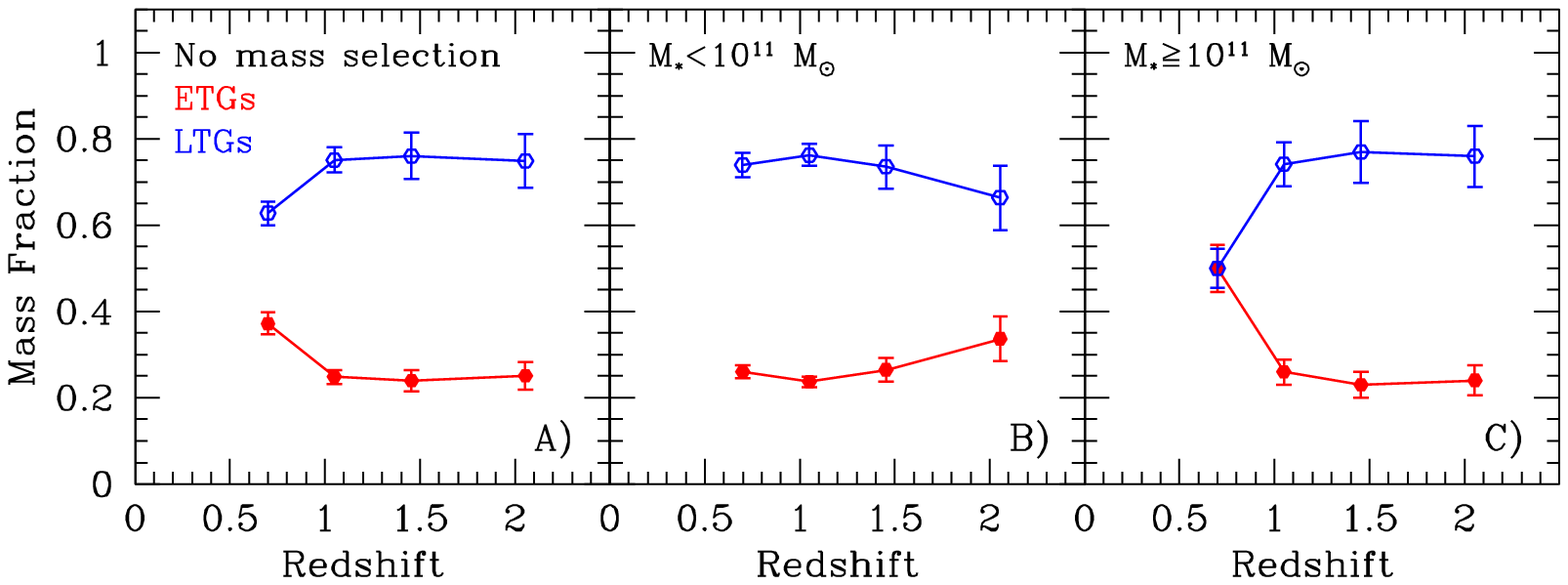}
\caption{Fraction of stellar mass in galaxies of different morphological type (ETGs in red, LTGs in blue) as a function of redshift for three bins of stellar mass. In panel A) the entire sample without mass selection is shown, in panels B) and C) the M$_*<10^{11}$ M$_\odot$ and the M$_*\geq10^{11}$ M$_\odot$ galaxies.}
\label{fig:mass_fraction}
\end{figure*}

Looking at the sample of massive galaxies shown in the right panel of Fig. \ref{fig:fra_num_den}, we note a deviation from the constant trend of the morphological fraction in the lowest redshift bin, with massive ETGs that increase their fraction from $\sim20$ \% at high $z$ to more than 40 \% at $z<1$. This is in agreement with the evidence found in the local Universe where the present-day massive galaxies population is dominated by early-type morphology objects (Baldry et al. \citeyear{baldry04}, Conselice \citeyear{conselice06}). 

The relative increase of massive ETGs appears qualitatively consistent with the findings of \citet{buitrago13} which can rely on a sample of about 700 galaxies in the range of redshift $0.6<z<2$.
They reported that the fraction of ellipticals have little or no evolution in the redshift range $1\lesssim$z$\lesssim2.5$ as we found, while at $z\lesssim1$ the ETGs strongly increase, becoming the predominant morphological class of massive galaxies. At $z\lesssim1$ \citet{buitrago13} detect higher fraction of ETGs compared to our work, but this can be due to their inclusion of lenticular galaxies in their ETGs sample.  

To trace the rate of growth of the two populations of galaxies and to quantify their evolution along the cosmic time, we estimated the comoving number density of ETGs and LTGs (bottom row of Fig. \ref{fig:fra_num_den}). We considered the same mass ranges and redshift bins of Fig. \ref{fig:fra_num_den} top row. The results show the number density for the total sample of 1302 galaxies (black) and for the two morphological samples of early- (red) and late-type galaxies (blue).
We see a large increase in the number density of the global population at decreasing redshift for all the mass ranges considered that reflects the rapid growth of the two morphological classes.
For the massive galaxies sample, complete in stellar mass, we computed the rate of growth of the number density fitting the data (Table 3) with a power law of the form $\rho_n=\rho_0\times(1+z)^{\alpha}$. The fit (solid lines of panel C) confirms the
significant growth of the total galaxy population that numerically increases by a factor of 8 from $z=2.5$ to $z=0.6$. The  growth is steeper for massive ETGs ($\alpha_{ETGs}=-4.23\pm0.61$) compared to a milder growth of the massive LTGs ($\alpha_{LTGs}=-3.35\pm0.43$). Indeed, the massive ETGs at $z\sim0.6$ are a factor of $\sim12$ more numerous than at $z\sim2-2.5$ (in agreement with Muzzin et al. \citeyear{muzzin13}), to be compared with a factor of $\sim7$ for the massive LTGs; the steepening of the $\rho_n$ is more pronounced in the lowest redshift bin where the two morphological classes became approximately equal in number.  
The effect of the rapid growth of the ETGs number density for $z<1$ could be explained by several processes including the formation of new massive ellipticals through the merging of less massive galaxies or by the morphological evolution of some massive discs towards elliptical morphologies.
The main implication of our finding for galaxies more massive than $10^{11} M_{\odot}$ is that irrespective of their morphology the majority of the massive galaxies are not yet in place until $z\sim0.6$. For redshift lower than z=0.8-1 several authors (Cimatti et al. \citeyear{cimatti06}, Ilbert et al. \citeyear{ilbert13}, Buitrago et al. \citeyear{buitrago13}) reported that the number density of massive ETGs is nearly constant and that no new massive ETGs formed within that redshift interval. We cannot exclude the eventuality that the trend of growth of the $\rho_n$ found up to z=0.6 tends to become constant at lower redshift, although in our lowest range of redshift (z=0.6-0.8) we do not observe any sign of flattening.

 \begin{table*}
 \label{tab:numden}
 \centering
 \begin{tabular}{llllllll}\hline
Redshift &N$_{gal}$ &$\rho_n$(tot) & $\rho_n$(ETGs) & $\rho_n$(LTGs) & $\rho_*$(tot) & $\rho_*$(ETGs) & $\rho_*$(LTGs)\\
& &Mpc$^{-3}$& Mpc$^{-3}$& Mpc$^{-3}$& M$_{\odot}$ Mpc$^{-3}$&M$_{\odot}$ Mpc$^{-3}$&M$_{\odot}$ Mpc$^{-3}$\\\hline\hline
$0.6\leq z\leq0.8$&43 &$98.0\pm15.0$& $41.0\pm9.7$ & $57.0\pm11.4$ &$16.4\pm1.0$ & $8.2\pm0.8$ & $8.2\pm0.6$  \\
$0.8<z\leq1.3$    &62 &$37.6\pm4.8$ & $7.9\pm2.2$  & $29.7\pm4.2$  &$7.5\pm0.3$  & $1.9\pm0.2$ & $5.5\pm0.3$  \\
$1.3<z\leq1.6$    &40 &$31.1\pm4.9$ & $8.6\pm2.6$  & $22.5\pm4.2$  &$7.0\pm0.4$  & $1.6\pm0.2$ & $5.4\pm0.4$  \\
$1.6<z\leq2.5$    &42 &$10.2\pm1.6$ & $2.9\pm0.8$  & $7.3\pm1.3$   &$2.7\pm0.2$  & $0.7\pm0.1$ & $2.1\pm0.1$  \\ \hline
 \end{tabular} 
 \caption{Comoving number densities (in $10^{-5}$ dex Mpc$^{-3}$ units) and mass densities (in $10^{7}$ M$_{\odot}$ Mpc$^{-3}$ units) for the complete sample of 187 massive galaxies (M$_*\geq10^{11}$ M$_\odot$) at $0.6\leq z\leq2.5$.}
 \end{table*}

\subsection{The stellar mass content of different morphological types}

In this section we present the study of the mass fraction and of the stellar mass density over the 5 Gyr of time that occurred in the redshift range $0.6\leq z\leq2.5$. 
\begin{figure}
\includegraphics[width=8.5 truecm]{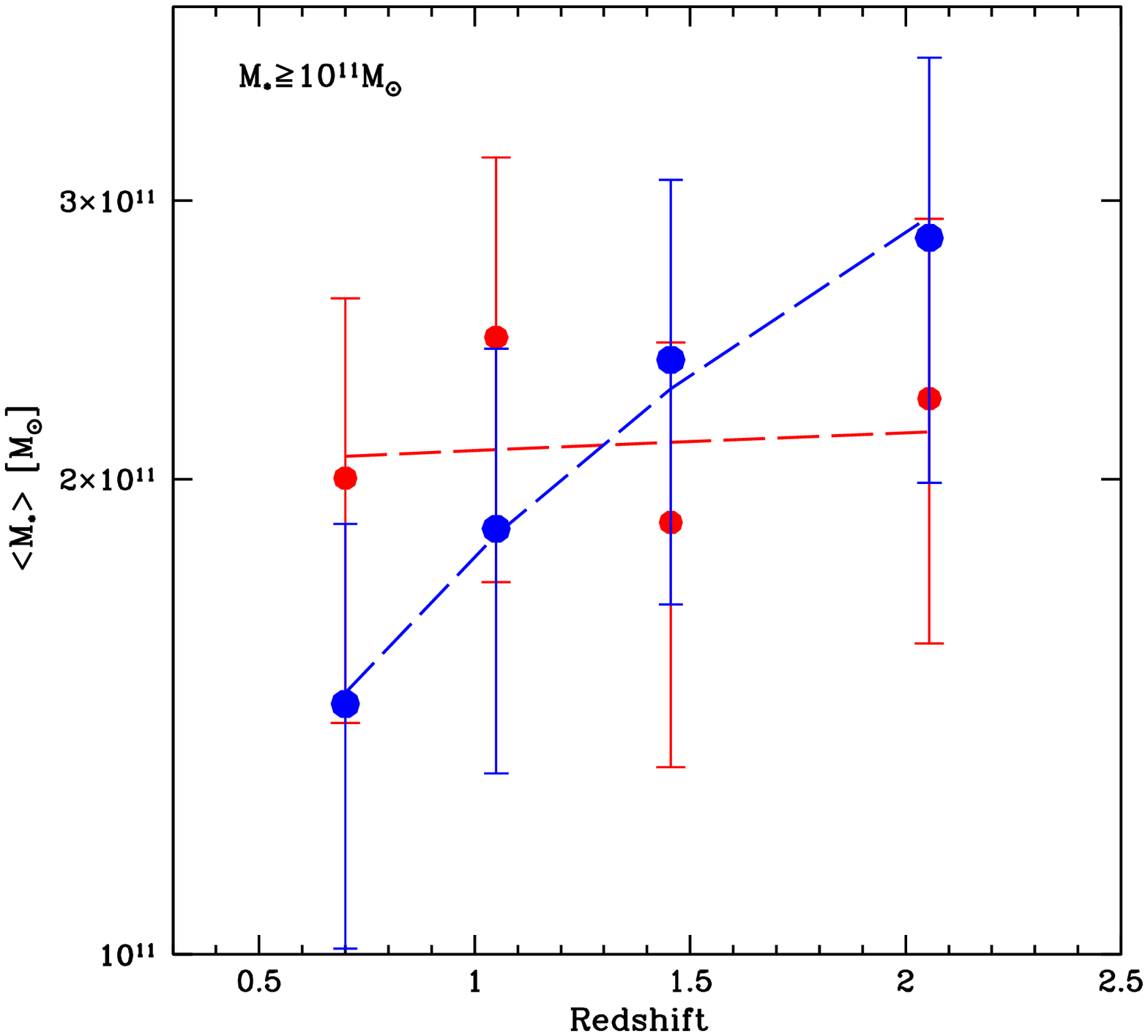}
\caption{Mean stellar mass as a function of redshift for the sample of galaxies more massive than M$_*\geq10^{11}$ M$_\odot$. Red and blue refer to the morphological classes of early- and late-type galaxies. The dotted lines represent a weighted least squares fit to the points. Error bars represent the typical uncertainty (30 \%) we assumed on the stellar masses.
}
\label{fig:massamedia}
\end{figure}
The stellar masses adopted in this analysis were estimated assuming a Salpeter IMF; an alternative choice of the IMF would cause a shift in the derived values of stellar masses without affecting the trends of our results.
For instance, masses obtained assuming a Salpeter IMF can be converted into those with Chabrier IMF dividing the former by a factor of 1.8 \citep{longhetti09}, the same applies to the other IMFs discussed in this paper.
As characteristic uncertainty on the stellar masses, we assume a typical error of 30 \%.

The stellar mass fraction as a function of redshift is shown in Fig. \ref{fig:mass_fraction}. We observe that the stellar mass fraction held in spheroidal galaxies tends to increase at $z<1$ (Fig. \ref{fig:mass_fraction} A). This growth disappears when we consider low-intermediate mass galaxies (M$_*<10^{11}$ M$_\odot$, Fig. \ref{fig:mass_fraction} B), for which we find the same constant trend found for the number fraction of the two morphological types over the whole redshift range considered. 
When we consider the massive galaxies (Fig. \ref{fig:mass_fraction} C) we find that the stellar mass contained in ETGs is constant at $1<z<2.5$, while it increases from $\sim25\%$ to $\sim50\%$ at $z<1$ reflecting roughly the number fraction growth. Thus, the increase in the mass fraction of ETGs observed at $z<1$ in the total sample (Fig. \ref{fig:mass_fraction} A) is due to the most massive spheroids. At $z\sim0.6$ the fraction of stellar mass locked in the massive spheroidal population is similar to that in massive disc-shaped and irregular galaxies. The apparent growth of the mass fraction locked in massive ETGs is also due to a decrease in the mass locked in massive late-type galaxies. Indeed, looking at Fig. \ref{fig:massamedia} the mean stellar mass contained in massive LTGs decreases with decreasing redshift, while it tends to be constant for the massive spheroidal population. 

\begin{figure*}
\includegraphics[scale=1]{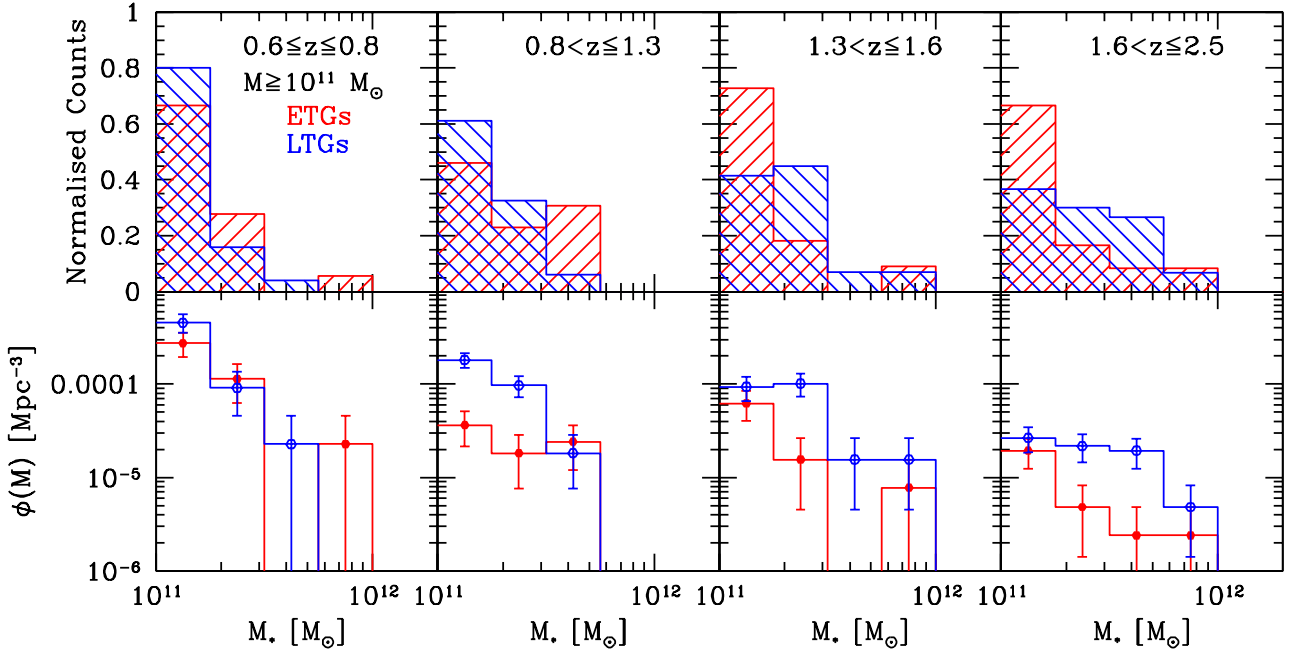}
\caption{Stellar mass (top panels) distributions and mass function distributions (bottom panels) of massive (M$_*\geq10^{11}$ M$_\odot$) ETGs (red) and massive LTGs (blue). The samples are split in the same redshift intervals adopted in Fig. \ref{fig:fra_num_den} and Fig. \ref{fig:mass_fraction}. The binning in stellar mass is not linear; it is performed between $log(\frac{M_*}{M_{\odot}})=11$ and $log(\frac{M_*}{M_{\odot}})=12$ with a step of 0.25.} 
\label{fig:massfunction}
\end{figure*}

In the top row of Fig. \ref{fig:massfunction} the normalised stellar mass distributions in four redshift bins for the  M$_*\geq10^{11}$ M$_\odot$ sample are shown. It can be seen that as the redshift decreases the distributions of the late-type population become more populated at the lower mass end (M$\sim10^{11}$ M$_\odot$), while for the massive ETGs the distributions do not change significantly over the redshift range covered by our sample. 
A similar trend is found in the distribution of the number density as a function of the stellar mass (Fig. \ref{fig:massfunction} bottom row), that is the stellar mass function of galaxies more massive than $10^{11}$ M$_\odot$.
For the late-type morphology as the redshift decreases we observe a clear trend to largely populate the low-mass end of the considered mass range and a tendency at the extremely-massive regime to remain steadily populated or even to empty at the highest mass bin. On the other hand, the early-type population (54 ETGs) shows extremely-massive galaxies both at high and at low redshift and the same tendency of LTGs to replenish the low-mass end distribution. 
Despite low statistics, the data show that the extremely-massive galaxies (M$_*>3-4\times10^{11}$ M$_\odot$) were already in place at $z\sim2$ and that their number did not increase substantially, otherwise we should observe some growth at lower redshift as happens for the less massive galaxies ($10^{11}\leq $M$_*$/M$_{\odot}\leq 3\times10^{11}$), hence the population of extremely massive galaxies formed at $z>2.5-3$.

If it is real and not due to statistical fluctuation, the lack of new extremely-massive LTGs at $z<1.3$ seen in Fig. \ref{fig:massfunction} could be evidence of a morphological transformation of these galaxies into more massive spheroids. However, we observe that the extremely massive LTGs tend to have high values of specific star formation rate and early formation epochs $z_{form}>5$. Thus a mechanism capable of shutting off their star formation and of removing the disc so that the morphological transformation occurs should take place in this hypothesis.

\begin{figure}
\includegraphics[width=8.5cm]{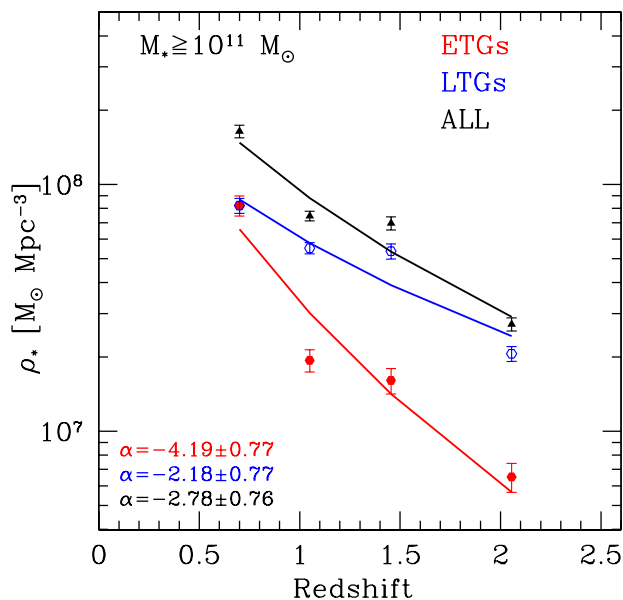}
\caption{Stellar mass density as a function of redshift, split by morphology, and with a mass cut of M$_*\geq10^{11}$ M$_{\odot}$. The solid lines illustrate power-law fits of the form $\rho_*=\rho_{*,0}\times(1+z)^{\alpha}$. }
\label{fig:mass_den}
\end{figure}

In Fig. \ref{fig:mass_den} the comoving stellar mass density as a function of redshift for the two samples of massive ETGs and LTGs is shown (data in Table 3). We observe an overall increase in the mass density, although the evolution for the massive LTGs is more modest compared to the one observed for massive ETGs.
Fitting our data with a power law form $\rho_*=\rho_{*,0}\times(1+z)^{\alpha}$ we found a significant difference between the slope of the ETGs and LTGs mass density with $\alpha_{ETGs}=-4.01\pm0.77$ and $\alpha_{LTGs}=-2.19\pm0.77$. This difference is mainly due to the steepening we observe for spheroidal galaxies between $z\sim1$ and $z\sim0.6$. By comparing the number and the mass density (Fig. \ref{fig:fra_num_den}c, second line, and Fig. \ref{fig:mass_den}) of the massive galaxies from $z\sim2$ to $z\sim0.6$, we note that both the number and the mass density of ETGs increase by an order of magnitude ($\sim11-12$), in agreement with \citet{brammer11}. On the contrary, the late-type population increases its number by a factor of $\sim7$ while its mass by a factor of $\sim3.5$.
This implies that the new massive late-type objects are progressively added with smaller masses, as shown before by the average mass contained in LTGs that declines by a factor of 2 since $z\sim2.5$.

The analysis of the number densities shows that the population of galaxies has significantly grown by an order of magnitude since z=2.5 regardless of its morphology and their mass. In particular, for the population of massive galaxies, combining the information from the number and the mass densities, we found that the galaxies with  M$_*>3-4\times10^{11}$ M$_\odot$ were likely formed at $z>2.5-3$ and that their assembly could not be effective at lower redshift. Moreover the massive LTGs at low redshift were formed with smaller masses compared to the massive ETGs.

The evolution of the number density cannot constrain a formation scenario of the massive ETGs and LTGs, since the data describes a snapshot in which is not possible drawing an evolutionary path.
We can just use the information contained in the mass function to claim that in the high-mass regime some morphological transformation may occur to justify the disappearing of extremely-massive LTGs.

\subsection{Dating the stellar mass content of massive galaxies}
In this section we investigate when the stellar mass content of the massive galaxies formed.
To this end, we estimated for each galaxy with M$_*\geq10^{11}$ M$_{\odot}$ the epoch $t_{form}$ at which the bulk of its stellar mass has been formed, 
\begin{equation}
 t_{form}=Age_{Univ}(z_{gal})-Age_{star}(z_{gal}),
\end{equation} 
where $Age_{Univ}(z_{gal})$ is the age of the Universe at the galaxy redshift and $Age_{star}(z_{gal})$ is the mean age of the galaxy stellar population at its redshift.

The age of the stellar population was derived by best fitting the galaxy SED with stellar population synthesis model as described in Sec. \ref{sec:passive} , hence it corresponds to the model age obtained assuming an exponentially declining star formation history. It represents the age of the stars dominating the overall galaxy emission and includes the contribution of stellar remnants.

\begin{figure*}
\includegraphics[width=8.5 truecm]{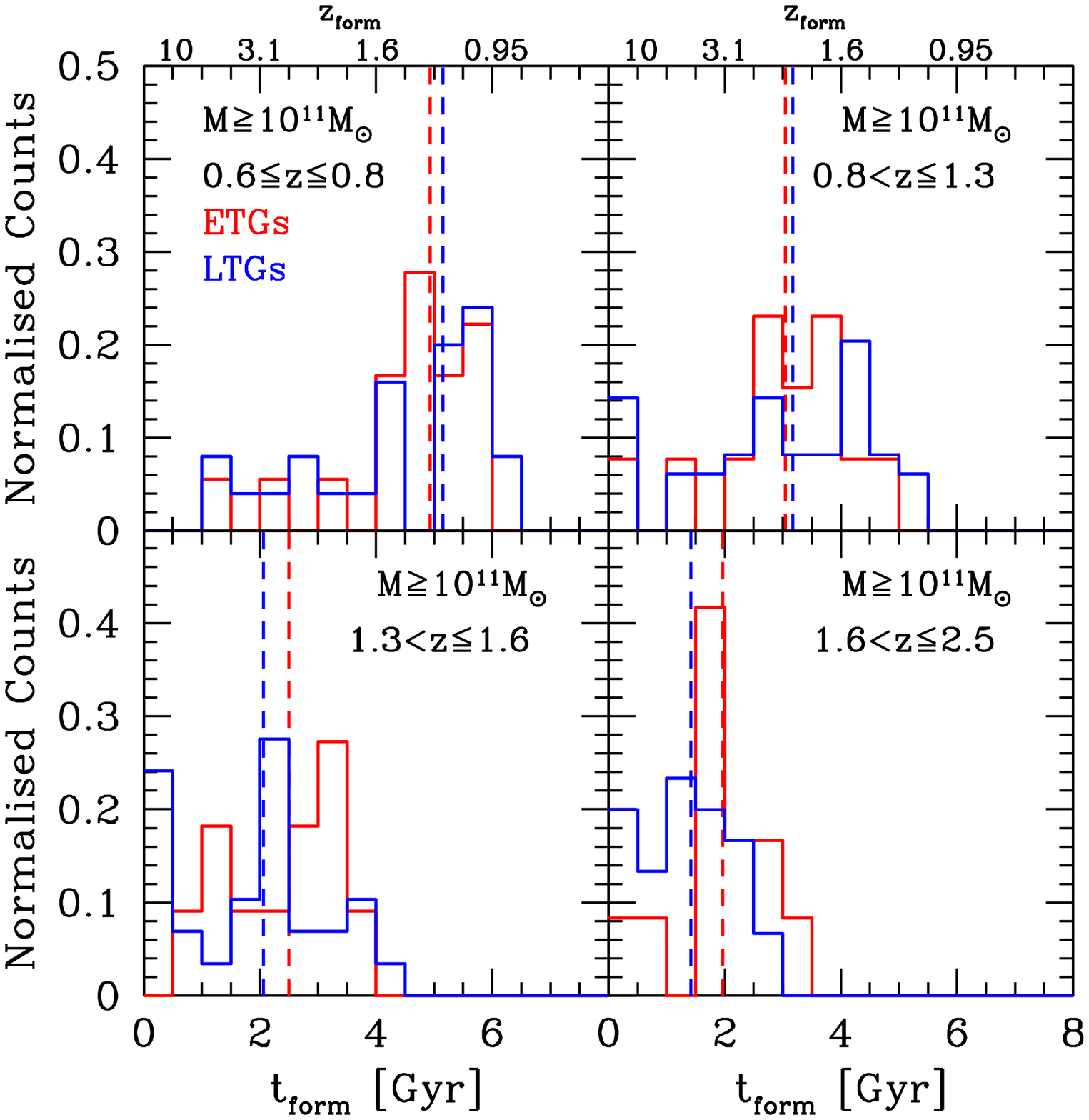}
\includegraphics[width=8.5 truecm]{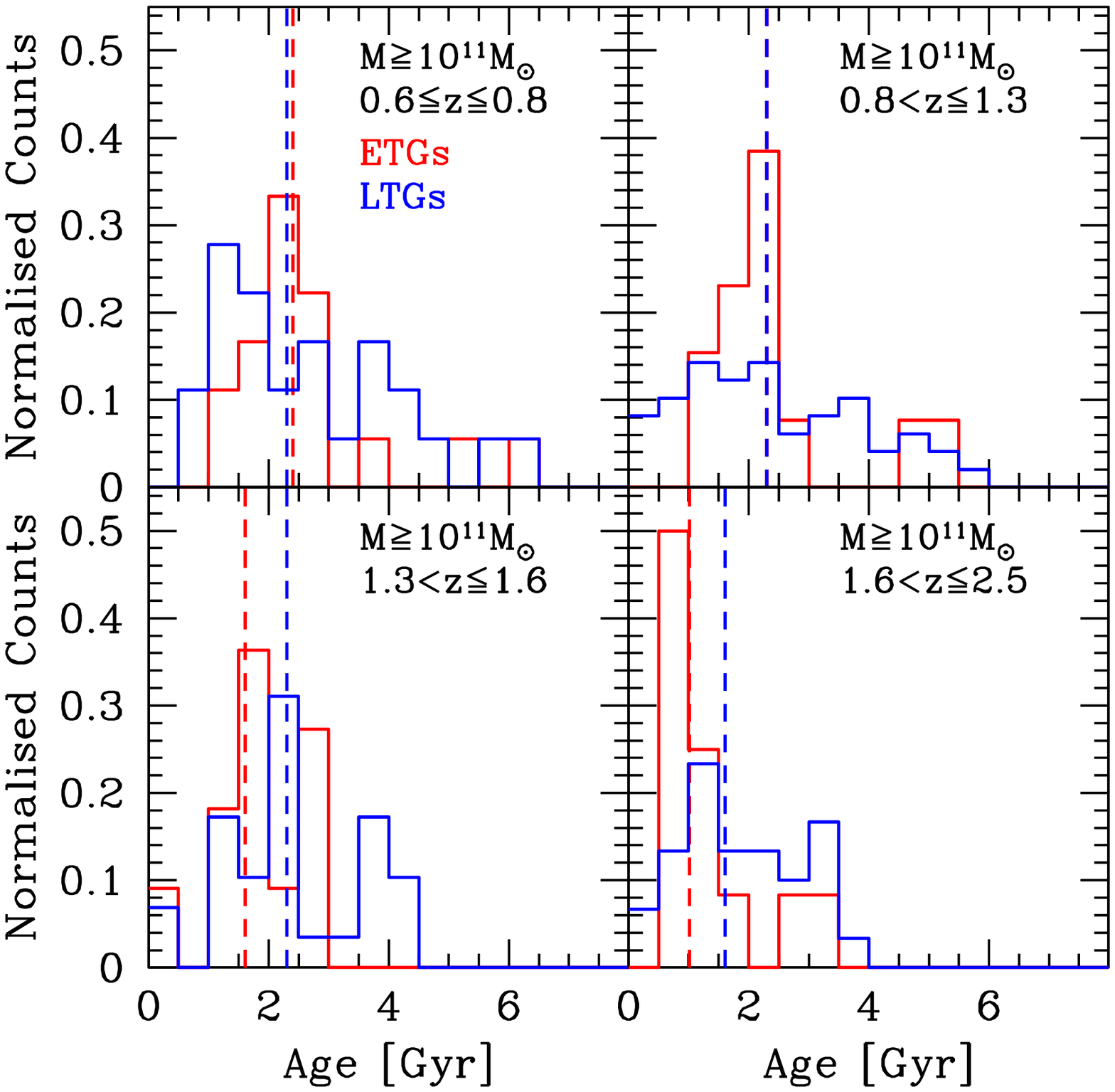}
\caption{Distributions of formation epochs (left panels) and ages (right panel) of the stellar component of massive galaxies (M$_*\geq10^{11}$ M$_\odot$) split into four redshift bins. Red and blue represent the ETGs and LTGs classes. Dotted vertical lines represent the median values. In the left panels the formation redshift $z_{form}$ ($z_{form}=z(t_{form}$) is indicated on the top x-axis.}
\label{fig:age_zbin}
\end{figure*}

In the left panels of Fig. \ref{fig:age_zbin} we show the normalised distributions of the formation epoch $t_{form}$ for early- (red) and late-type (blue) galaxies in four different redshift bins. It is clearly visible that the formation redshift $z_{form}=z(t_{form})$ of the stellar population of massive galaxies are spread over a wide cosmic time that varies between $z_{form}\sim1$ and $z_{form}\sim10$, without a preferred time for the formation of the bulk of the stars. The distributions suggest that the majority of galaxies apparently formed their stars at an epoch decreasing with decreasing  redshift, with the bulk of stellar mass produced within a few Gyrs of the observation redshift. This result is not surprising if we consider how the number density of massive galaxies evolves (Fig. \ref{fig:fra_num_den}, bottom right). Indeed we have seen that new massive galaxies are progressively added as redshift decreases, thus only a small fraction of galaxies observed in the lower redshift bins were already assembled at higher redshift. In each redshift bin we observe a tail of galaxies with stellar populations formed at earlier epochs. However, the bulk of the stars in each redshift bin was created when the Universe was a few Gyrs younger than the time at which we observe them. In the right panels of Fig. \ref{fig:age_zbin} the distributions of the mean ages of stars contained in massive galaxies are shown; we note the same wide spread observed for the formation epochs with median ages of the stellar components of about 2 Gyr for every redshift bin and morphological type.

Summarising, we have found that the galaxy number density grows with decreasing redshift, thus new massive galaxies are formed as the cosmic time passes. Moreover, the age of the stellar component of these galaxies in each redshift bin is generally 1-2 Gyr older than the observation time. Thus, the epoch at which stars have been formed and the time at which the galaxy has been assembled could not be temporally distant.
Alternatively massive ETGs, at least some of them, after a primary initial episode of star formation have increased their mass through further star formation events that lower the mean age of the resulting stellar population.
Another possibility could be the merging between subunits with a younger stellar population occurred at different times or a star formation feeding by cold streams and minor mergers (Birnboim \& Dekel \citeyear{birnboim03}, Kere\v{s} et al. \citeyear{keres05}, Ceverino et al. \citeyear{ceverino12}) that still shift the formation epoch to a more recent time. 

We emphasize that for the age of the stellar populations we use the age of the model best fitting the observed SED. This age cannot keep track of the presence of a more complex composition of the stellar populations inside the galaxies. For instance, \citet{lonoce14} through the analysis of spectroscopic indices of a sample of 15 ETGs at $0.7<z<1.1$ have found that the main component ages tend to be older than the ages derived by the standard SED fitting. 
As known, the optical blue and UV ranges of a galaxy SED are typically dominated by very luminous young stellar populations even when they represent a small percentage of the total stellar content of the galaxies. Since our sample covers the range from z=0.6 to z=2.5, at low redshift we do not have the same good sampling of the far-UV wavelengths (rest frame) as happens for higher redshift. Thus, the ages measured in the range z=0.6-0.8 are affected by a poor SED coverage in the far-UV range. 
With the aim to evaluate the dependence of age estimates on the rest-frame wavelength coverage, we repeated the SED-fitting process to ensure that all the galaxies had the same rest frame photometric coverage. We performed a new SED-fitting eliminating from the fitting process those filters bluer than the outermost UV$_{rest}$  filter at z=0.6 ($\lambda_{r. f.}\sim 2250$ $\AA{}$).
We found that shortening the UV$_{rest}$ coverage implies slightly younger ages. Therefore, if we had in the bin $z=0.6-0.8$ all the coverage available at higher redshift we would probably have obtained even younger ages. This result guarantees that the recent formation epochs found for the bulk of the stellar populations at $0.6\leq z \leq0.8$ are not due to the filter choice and that the behaviour we found would not change with different wavelength coverage.

\section{Summary and conclusions}
Starting from a sample of 1302 galaxies, limited at Ks$\leq22$, with spectroscopic coverage of 76 \% selected in the GOODS-South field, we have selected a sample of passive galaxies on the basis of their sSFR and a sample of early-type galaxies classified on the basis of their morphology, to compare how the two selection criteria differ in terms of selected galaxies and main sample properties.

The morphological analysis that allowed us to select 247 ETGs (i.e. elliptical and spheroidal galaxies), was carried out on ACS-F850LP images and was also supported by inspections of residual maps resulting from the luminosity profile fitting with S\'{e}rsic models.
The passive galaxy samples were selected on the basis of a specific star formation rate sSFR=SFR/M$_{*}\leq10^{11}$ yr$^{-1}$, where stellar mass and star formation rate were estimated by fitting the observed SED with stellar population synthesis models that assumed different IMFs. The Initial Mass Functions adopted in the models have a Salpeter-like functional form that account for a different abundance of low-to-high mass stars.
Taking as reference the passive sample derived assuming the Salpeter IMF we study its difference with the morphologically selected sample of ETGs.
We found that $\sim35$ \% of passive galaxies have a late-type morphology and that about 26 \% of spheroids were missed by this selection. We find no significant differences in stellar masses, ages, and radii between the sample of passive galaxies and that of morphological ETGs. When we focused on the passive sample composed by galaxies with different morphologies we found that the main differences were observed on the axial ratio distributions of passive ETGs and passive LTGs, stressing their different morphology.

We find that the passive samples and their morphological mix strongly depends on the IMF assumed in the spectrophotometric models. Indeed, increasing the abundance of low-mass stars in the IMF the number of classified passive galaxies decreases. Simultaneously, the number of ETGs collected and the relative contamination of late-type objects decrease. 

We used the visually classified sample of 196 ETGs compared with the broad class of 723 LTGs to follow their evolution over the redshift range $0.6\leq z\leq2.5$. The main results of our analysis can be summarised as follows.\\

 (i) The relative fraction of ETGs and LTGs is constant over the redshift range considered, with spheroidal galaxies settled at 20-30 \% and LTGs at 70-80 \%. Hence, the population of galaxies does not change significantly their composition across the time at least for masses $<10^{11}$ M$_{\odot}$. For galaxies with M$_{*}\geq10^{11}$ M$_{\odot}$ at $z<1$ we found a significant change in the composition of the morphological fraction; in particular we observe an increase in the fraction of massive ETGs to more than 40 \% of the massive population. These results suggest that the Hubble sequence was already in place more than 10 Gyr ago.\\

 (ii) We also investigate the number density evolution of the two classes, finding that irrespective of their morphology and mass, both spheroids and discs have significantly increased their number density from z=2.5 to z=0.6, with LTGs being always the richer population. For the complete sample of massive galaxies (M$_{*}\geq10^{11}$ M$_{\odot}$), we found that between $z=2.5$ and $z=0.6$ the whole population of galaxies increased by a factor of $\sim8$. This growth was faster for the massive ETGs population whose number density increases by a factor of $\sim$12, than for the massive LTGs that increased by a factor of $\sim7$. This difference, however, is mainly due to the steepening we observe for spheroidal galaxies between $z\sim1$ and $z\sim0.6$.\\

 (iii) We examined the mass fraction evolution for different stellar mass bins and we found that for the low-intermediate mass regime (M$_{*}<10^{11}$ M$_{\odot}$) the mass fractions follow the same constant evolution, with similar percentage of the number fractions. For the massive ETGs we observe an increase in the mass fraction toward lower redshift and a growth in the mass density with a similar rate of the number densities. The rise of the mass fraction of massive ETGs is due to a drop of the stellar mass assembled at lower redshift in massive discs. We found that the rate of growth of the mass density in massive LTGs is halved compared to the increase in their number density.\\

 (iv) Our data show that the extremely-massive galaxies (M$_{*}\gtrsim4\times10^{11}$ M$_{\odot}$) were in place at high redshift ($z>2.5$) and that their number did not increase as the less massive ones did ($10^{11}<$M$_{*}<4\times10^{11}$ M$_{\odot}$), but remained approximately constant, suggesting a scenario in which extremely-massive galaxies were mostly formed in the early Universe. Moreover, as the redshift decreases we observed a deviation toward the low-mass end of the stellar mass distribution of the massive LTGs, also recovered in their mass function distribution. The presence of extremely-massive LTGs only at higher redshift suggests a disc-to-spheroid transformation, through one or more processes that quenches the star formation and remove the disc such as the dynamical instabilities mechanism (Kormendy \& Kennicut 2004).\\

 (v) Finally we found that the formation epochs of massive galaxies span a broad range of time with no particular indication towards a privileged time for star formation and with mean ages of a few Gyr younger than the observation redshift. This is not surprising considering the rise of the number density and suggests that new massive galaxies both ETGs and LTGs are assembled as the time flows, with stellar populations of a few Gyr old.

\section*{Acknowledgements}
This work has received financial support from Prin-INAF 1.05.09.01.05.

\nocite{}
\bibliographystyle{aa}
\bibliography{paper_tamburri}

\appendix
\section{Structural parameters}
We performed the luminosity profile analysis, to determine the morphological parameters of the galaxies, on the HST-ACS images of the GOODS-S field in the F850LP filter. These data provide high signal-to-noise ratio, images with good spatial resolution (FWHM$\sim$0.11'' measured on the images) and an excellent sampling of the PSF (pixel size=0.05''). 

We decided to fit the surface brightness (SB) profile of the observed galaxies with a single S\'{e}rsic law,
\begin{equation}
 \mu (r)= \mu_e + \frac{2.5 b_n}{ln(10)}[{(\frac{r}{r_e})}^{1/n}-1],
\end{equation} 
where $\mu(r)$ is the SB measure at distance $r$ from the centre of the galaxy, $r_e$ is the effective radii of the galaxies containing half of the total light, $\mu_e$ is the normalisation term of the SB when $r=r_e$, $n$ is the S\'{e}rsic index, and $b_n$ is a parameter depending on $n$ ($b_n\sim2n-1/3+0.009876/n$, Prugniel \& Simien \citeyear{prugniel97}).
When the S\'{e}rsic index $n$ is large, the profile is steep for small radii and highly extended in the outer regions; otherwise for small $n$, the profile has a flatter core with a sharp truncation at large radii.
 
The fit of the light profile was performed with the \texttt{GALFIT} software \citep[v. 3.0.4;][]{peng02}, a routine that provides the best fitting parameters of the profile, minimizing the differences (residuals) between the PSF-convolved model and the input galaxy images. From the fit procedure we obtained the axial ratio $b/a$, the  S\'{e}rsic index $n$, and the semi-major axis $a_e$. The circularised effective radius $r_e$ is related to $a_e$ through the equation $r_e=a_e\sqrt{b/a}$.

The fitting process was repeated four times for each galaxy, using four different PSFs of unsaturated and isolated stars, in order to exclude the dependence of the fitting results from the PSF used. As best fitting parameters we chose the one with the minimum chi-squared.
We used as input a sigma image, i.e. the image showing pixel-to-pixel flux uncertainty, the rms map produced by Sextractor using the standard HST-ACS weighting image. 
We also tried to generate an internal \texttt{GALFIT} sigma image and we verified that the fitting results did not significantly change compared to the results obtained using the rms map produced with Sextractor. The fitting was performed over a box of 6 x 6 arcsec (201x201 pixel) centred on the centroid of each galaxy, except for galaxies with highly extended structures for which a larger box has been adopted. The convolution box size and the PSF image size were fixed at 3 x 3 arcsec (101x101 pixel).
To define the size of the fitting box we applied an iterative technique, consisting in enlarging the box size around the target source until the parameters converge with growing fitting area. This process was repeated also to fix the convolution and the PSF box size and to ensure that the structural parameters obtained with \texttt{GALFIT} were not biased by the choice of these input parameters. 
Every object close to a target was fitted simultaneously, while more distant and weak sources were masked out using the Sextractor segmentation image, these processes prevent the profile of the surrounding galaxies from overlapping the target profile.
Bad fitting results, i.e. fits that do not converge or those for which one or more parameters have been forced to a fixed value in \texttt{GALFIT}, are around 5\%.

\end{document}